\documentclass[12pt]{article}

\usepackage{amsmath}
\usepackage{amsthm}
\usepackage{amsfonts}          % if you want the fonts
\usepackage{amssymb}           % if you want extra symbols
\usepackage[all]{xy}           % Commutative diagrams
\usepackage{epsfig}

\newlength{\xtrawidth}
\newlength{\xtraheight}
\newcommand{\eref}[1]{eq.~\eqref{#1}}
\newcommand{\sref}[1]{\S\ref{#1}}

\newcommand{\fref}[1]{Fig.~\ref{#1}}
\newcommand{\cref}[1]{Chapter~\ref{#1}}
\newcommand{\beq}{\begin{equation}}
\newcommand{\eeq}{\end{equation}}
\newcommand{\bea}{\begin{eqnarray}}
\newcommand{\eea}{\end{eqnarray}}
\newcommand{\bean}{\begin{eqnarray*}}
\newcommand{\eean}{\end{eqnarray*}}
\newcommand{\ba}{\begin{array}}
\newcommand{\ea}{\end{array}}

\newcommand{\tr}{\ensuremath{\mbox{Tr}}}

\def\IC{\mathbb{C}}
\def\II{\mathbb{I}}

\def\IR{\mathbb{R}}

\def\IZ{\mathbb{Z}}

\def\IP{\mathbb{P}}

\def\cN{{\mathcal N}}

\def\cO{{\mathcal O}}
\def\cC{{\mathcal C}}

\def\W{\widetilde}

\newcommand{\mat}[1]{\left( \begin{matrix}#1\end{matrix} \right)}
\newcommand{\qq}[1]{\begin{quote}{\sf #1}\end{quote}}

\def\nn{\nonumber}

\newtheorem{theorem}{\bf THEOREM}

\newcommand{\setall}{\setcounter{equation}{0}
        \setcounter{theorem}{0}}

\newcommand{\gen}[1]{ \langle #1 \rangle}
\newcommand{\comment}[1]{}

\begin{document}
\rightline{\small MIT-CTP 3761}
\rightline{\small CMSZJU-06005}
%\rightline{\small ????}

\vskip 0.5in
\centerline{\LARGE \bf Counting BPS Operators in Gauge Theories}
{\flushright{\Large {\bf - Quivers, Syzygies and Plethystics\\}}}
~\\

\renewcommand{\thefootnote}{\fnsymbol{footnote}}

\centerline{{\bf 
Sergio Benvenuti${}^{1,2}$\footnote{\tt s.benvenuti@sns.it}, 
Bo Feng${}^{3,4}$\footnote{\tt b.feng@imperial.ac.uk} 
Amihay Hanany${}^{1}$\footnote{\tt hanany@mit.edu}, 
Yang-Hui He${}^{5,6}$\footnote{\tt hey@maths.ox.ac.uk}, }}
{\small
\begin{flushleft}
${}^1${\it Center for Theoretical Physics, M.I.T.,
Cambridge, MA02139, U.S.A.\\}
${}^2${\it Joseph Henry Laoratories., Princeton University, Princeton, NJ.,
    U.S.A.\\}
${}^3${\it Blackett Lab.~\&
Inst.~for Math.~Science, Imperial College, London, SW7 2AZ, U.K.}\\
${}^4${\it Center of Mathematical Science, Zhejiang
University, Hangzhou 310027, P.~R.~China}\\
${}^5${\it Merton College,
Oxford University, Merton Street, Oxford, OX1 4JD, U.K.} \\
${}^6${\it Mathematical
Institute, Oxford University, 24-29 St.\ Giles', Oxford, OX1 3LB,
U.K.}
\end{flushleft}
}

\setcounter{footnote}{0}
\renewcommand{\thefootnote}{\arabic{footnote}}
\vskip 0.8in

\begin{abstract}
We develop a systematic and efficient
method of counting single-trace and
multi-trace BPS operators with two supercharges,
for world-volume gauge theories of $N$ D-brane probes 
for both $N \to \infty$ and finite $N$.
The techniques are applicable to generic
singularities, orbifold, toric, non-toric, complete intersections, 
et cetera, even to
geometries whose precise field theory duals are not yet known.
The so-called ``Plethystic Exponential'' provides a simple
bridge between (1) the defining equation of the Calabi-Yau, 
(2) the generating function of
single-trace BPS operators and (3) the 
generating function of multi-trace operators.
Mathematically, fascinating and intricate
inter-relations between gauge theory, algebraic geometry, combinatorics and
number theory exhibit themselves in the form of plethystics and syzygies.
\end{abstract}

\newpage

\tableofcontents

\vspace{1in}

%%
%-------------------INTRO
%%
\section{Introduction}
\setall
The study of BPS states in a quantum field theory is of unquestionable
importance.
The purpose of this note is to discuss the set of all mesonic BPS
gauge invariant operators (GIO) with two supercharges
which appear in the chiral ring of a generic $\cN=1$
supersymmetric gauge theory that lives on a D3-brane which probes a
singular Calabi Yau (CY) manifold\footnote{Some preliminary results
  were annouced in \cite{han-talk}.}. 
For arbitrary singularities,
finding the gauge theory living on the
D3-brane is intricate. The simplest class is the orbifolds, the study
of which began with
\cite{DM,Johnson:1996py,Kachru:1998ys,Lawrence:1998ja,Hanany:1998sd,Hanany:1999sp,Douglas:1997de}.
The next simplest class is the toric singularities, the investigation
of which was initiated by
\cite{Douglas:1997de,Morrison:1998cs,Beasley:1999uz,Feng:2000mi}. 
Interesting duality
structures of these theories have been expounded in
\cite{Feng:2001bn,Beasley:2001zp,Cachazo:2001sg,duality}.
It is recently realised that the toric theories are, in fact,
best described using a bi-partite periodic tiling of the
two dimensional plane, a so-called ``dimer model''
\cite{Hanany:2005ve,Franco:2005rj,Franco:2005sm,Butti:2005sw,Butti:2005vn,Hanany:2005ss,Feng:2005gw,Franco:2006gc,Hanany:2006nm,Garcia-Etxebarria:2006aq,Imamura:2006ub,Butti:2006nk,Benvenuti:2005cz}
(also cf.~\cite{Ueda:2006wy,Ueda:2006jn} for recent mathematical
treatments).

When the manifold is non-orbifold and non-toric
there is no current systematic way of describing the gauge theory even
though some examples exist in the literature. For example, the
higher del Pezzo series \cite{highdP} and certain deformations of
toric singularities \cite{Butti:2006nk} have been addressed. 
In this
paper we will see how one can describe the single- and
multi-trace operators in
terms of generating functions which can be computed for both toric and
non-toric manifolds. In fact, the computations we will see can
calculate the generating functions even for cases in which the gauge
theory is not precisely known - either the superpotential is missing
or even the quiver itself is not known.

The discussion on GIO's in the chiral ring can be
divided into few parts as follows. Given a gauge theory description of
the theory on the D-brane, there are several problems of interest:

\paragraph{Global $U(1)$ Charges: }
One
would like first to identify the set of global $U(1)$ charges of this
theory. One charge out of this set is singled out to be the {\bf
  R-charge}
and the other charges can be generically called global non-R
charges. The most useful way of thinking about these charges is by
introducing the holographically dual gravity description. A set of
D3-branes on a singular conical CY is holographically dual to an
$AdS_5\times Y_5$ background where $Y_5$ is a Sasaki Einstein (SE)
manifold (cf.~\cite{Martelli:2005tp} and references therein).
The global charges of the gauge theory are dual to gauge
fields in $AdS_5$. These gauge fields can be divided into two sets --
one set originates from the Type IIB metric, those are typically
referred to as the isometry of the SE manifold, and the other set
comes from the Type IIB 4-form. The R-charge is always part of the
isometry group of the SE manifold. The traditional name which was
given to the charges coming from the metric are {\bf flavor charges}
and those which come from the 4-form are called {\bf baryonic charges}.

The
isometry group of the SE manifold has a maximum rank of 3, in which
case the SE manifold and its CY cone are called {\it toric};  
the minimum rank is 1, in which case the corresponding $U(1)$ charge
is precisely the
R-charge. The number of baryonic charges is in principle unbounded and
is given by the third homology of the SE manifold. Most cases which were
studied in the literature have one baryonic symmetry, the prototypical
example being the conifold \cite{Klebanov:1998hh}. 
Currently there are extensive studies of
cases with more than one baryonic charge, the simplest being
the Suspended Pinch Point (SPP)
\cite{Morrison:1998cs,Franco:2005sm,Park:1999ep}, as well as the more
complicated $X^{p,q}$ family \cite{Benvenuti:2004wx}.

\paragraph{Counting Gauge Invariant Operators: }
Given the set of $U(1)$ symmetries, R, flavor, baryonic, etc., say, $n$
of them, each gauge invariant operator in the chiral ring carries a
set of charges under these symmetries. We will assign a generic
complex variable $t_i, i=1\ldots n$ to each such charge and define a
function $f(\{t_i\})$ to be the {\bf generating function} of all these
operators. This function $f$ has, by definition, an expansion in terms
of monomials in $\{t_i\}$ such that the coefficient, $c_{k_1, \ldots,
  k_n}$ of $t_{1}^{k_1}\cdots t_{n}^{k_n}$ is integer and counts the
number of operators of charges $(k_1,\ldots,k_n)$,
\beq
f(\{t_i\})= \sum\limits_{i_1,\ldots, i_k}
c_{k_1, \ldots,k_n} t_1^{k_1}\ldots t_n^{k_n} \ .
\eeq
Our goal is to compute such functions for a multitude of cases.

Our ultimate wish is that for any CY manifold we would like to know
\begin{enumerate}
\item The set of single-trace BPS operators, the generating function
  is denoted by $f$;
\item The set of multi-trace BPS operators, the generating function
  is denoted by $g$;
\item For $N$ D3-branes at the singular CY we would like to know the
  dependence on $N$. Namely, we would like to know how many
  independent single-trace and multi-trace operators are there in the
  chiral ring for a given set of charges. For a finite $N$ this turns
  out to be a much more 
  difficult task since there are matrix relations for a
  finite size matrix that need to be taken into account. Nevertheless,
  we propose a nice solution to this as well; the generating functions in
  this case will be denoted as $f_N$ for single-trace and $g_N$ for
  multi-trace.
\end{enumerate}

As will be discussed in detail in later sections there is an important
function which beautifully relates the single-trace generating
function and the multi-trace.
Namely, $g_N$ can be simply computed from $f_N$ using the
so-called {\it ``Plethystic Exponential."} 
This function has been used
in physics several times in the past and we believe it should go into
the literature more often as it plays a crucial role in counting
problems such as the one dealt with here\footnote{A.~H.~would like to
  thank Marcos Mari\~no for demonstrating the properties of this
  function \cite{Labastida:2001ts}.}.
Conversely one can use the so-called {\it ``Plethystic Logarithm"}
which is
the inverse function to the plethystic exponential and computes $f_N$
from $g_N$. The ability to switch between $f$ and $g$ will turn out to
be a key tool in analyzing the theories we are interested in and to
reveal new pieces of information which were either previously unknown
or alternatively not well discussed.

Having presented a host of functions and concepts, it would be most
expedient to exemplify them in a context with which the readers are
well-acquainted. We shall do so for the famous D3-brane theory on
$\IC^3$ in the next section. Having whetted the readers' appetites,
the plan for the remainder of the paper is as follows. We begin with
the large $N$ limit and present the solution to questions (1) and (2)
above. In \sref{s:single}, we show how to construct $f$, the
generating function for single-trace GIO's. This is a
Hilbert-Poincar\'e counting problem.
We exemplify with orbifolds, toric varieties and the del Pezzo
family. We take an interlude in \sref{s:dimer} and
examine this counting problem
using the graphical perspectives of dimers. Then, in \sref{s:multi},
we construct the generating function $g$, which count the multi-trace
GIO's. The relationship between $f$ and $g$ will turn out to be a
plethystic one. In due course, we will show how plethystics actually
encode not only the GIO's counting, but also the defining equation of the
singularity. Interesting partition identities as well as syzygies in
graded polynomial rings emerge. Having constructed the generating
functions, we then calculate the asymptotic behaviour thereof in
\sref{s:asym}, using results from combinatorics and analytic number
theory. Finally, we use the above formalism to address the more
difficult problem of finite $N$ in \sref{s:finiteN}
and show how plethstics again solves
the counting problem and how they encode the geometry. We conclude
with perspectives in \sref{s:conc}.

\section{$\IC^3$: An Illustrative Example}\label{s:c3eg}
\setall
As promised in the introduction, we begin with a familiar example to
illustrate the various generating functions.
Here, the computation can be done without
using the more general techniques which will follow in the rest of the
paper.
This example is of course for the archetypal example of the AdS/CFT
correspondence, the case in which the CY
manifold is trivially $\IC^3$ and its associated SE manifold, $S^5$
\cite{Maldacena:1997re}.
There are no baryonic charges
in this case since the third homology of $S^5$ is trivial and the
isometry group is $SU(4)$ with rank 3, meaning that this CY manifold is
actually toric and the number of $U(1)$ charges is 3.
We can thus define 3 corresponding
variables, $t_1, t_2, t_3$, which will then measure these three $U(1)$
charges in their powers, as explained above. The gauge theory is the
$\cN=4$ gauge theory with $U(N)$ gauge group which in $\cN=1$ language
has 3 adjoint chiral multiplets which we will denote as $x, y$ and
$z$. Being toric, this CY manifold admits a description in terms of
periodic bi-partite tilings of the two dimensional plane and in fact
is given by the simplest of them all - tilings by regular hexagons
\cite{Franco:2005rj}.

We are interested in operators in the chiral ring and therefore we
need to impose the F-term relations coming from the superpotential
$W = \tr(x[y,z])$. The F-terms hence take a particularly simple form:
$[x,y]=[y,z]=[z,x]=0$, i.e., all chiral adjoint fields commute.
The generic single-trace GIO in the chiral ring will then take
the form of $\tr(x^iy^jz^k)$. It is then natural to assign $t_1$ as
counting the number of $x$ fields, $t_2$, the number of $y$ fields and
$t_3$, the number of $z$ fields in a GIO. There will therefore be a
corresponding monomial $t_1^i t_2^j t_3^k$ for each gauge invariant of
charges $i, j,$ and $k$, respectively. In fact, there will be
precisely one for each triple of charges, provided each of $i,j$, and
$k$ are non-negative. Putting all of this together, we find that the
generating function $f$ takes the form
\beq\label{C3-f-ijk}
f(t_1,t_2,t_3; \IC^3)=\sum_{i=0}^\infty\sum_{j=0}^\infty\sum_{k=0}^\infty
t_1^i t_2^j t_3^k = \frac{1}{(1-t_1)(1-t_2)(1-t_3)} \ .
\eeq
To be more precise, in the above form we did not take into account any
relations that a finite matrix should satisfy, therefore, as
mentioned earlier, this result
is strictly valid for the case of $N=\infty$. Therefore, using the
notation introduced above we should write
\beq\label{C3}
f_\infty(t_1,t_2,t_3; \IC^3)= \frac{1}{(1-t_1)(1-t_2)(1-t_3)}.
\eeq

\paragraph{A General Feature for Toric CY: }
Note that in \eref{C3} the coefficients $c_{ijk}$ appearing in the
general expansion 
\beq
f_\infty(t_1,t_2,t_3) = \sum_{ijk} c_{ijk} t_1^i t_2^j t_3^k
\eeq 
are all equal to either 1 or 0. This means
that for a given set of charges, $i,j,k$, there is either one
operator carrying these charges or not, but there can not be more
than one. Indeed this is a generic feature which is obeyed for every
toric singular CY. More explicitly there is a one-to-one
correspondence between single-trace GIOs and integer lattice points
in the dual cone of toric diagram
\cite{Beasley:1999uz,Butti:2006nk,Martelli:2006yb}.
This property is reminiscent of
some kind of a fermionic degree of freedom that carries this set of
charges. In contrast, for the non-toric case, it is shown in
\cite{Butti:2006nk} that there are, in general, multiple-to-one mappings
between single-trace GIOs and given charges. The reason is clear. In
the toric case, we have two extra $U(1)$ flavor symmetries besides
the R-symmetry, which is big enough to distinguish finely,
while for non-toric case we do not have these extra symmetries.

Let us look at the set of charges $i,j,k$ for which $c_{ijk}$ are
not zero. They form a sub-lattice of the three dimensional lattice
which has the form of a cone. Indeed, this sub-lattice is the 
so-called ``positive octant" for which $i\ge0, j\ge 0, k\ge0$. This
feature of a cone structure will also be general for every CY
manifold, the form of this cone is interesting and will be discussed
in detail in \sref{s:dimer}. One can think of the function $f_\infty$
as a theta-function over the lattice points of the cone and is a
characteristic function of this cone. It is worth to notice that
from results in \cite{Butti:2006nk}, it seems that there is a lattice
structure for both toric and non-toric cases. The difference is
that for the toric case the lattice is 3-dimensional 
while for non-toric case the dimension is lower.

If on the other hand we are interested in counting the number of BPS
operators which carry a given fixed scaling dimension, say
$\tr(x^iy^jz^k)$ of dimension $i+j+k = \frac{3}{2}R$, we need to
set $t_1=t_2=t_3=t$ in \eref{C3} and get the generating function for
all operators. In other words, we have to forget the other two $U(1)$
flavor symmetries and use the fact that all variables $x,y,z$
have same R-charge ${2\over 3}$. Hence,
\beq 
f_\infty(t;~\IC^3) = \frac{1}{(1-t)^3} = \sum_{m=0}^\infty {m+2
\choose 2}t^m
\eeq 
and the number of GIO's of given
R-charge $R=\frac{2}{3}m$ is ${m+2 \choose 2}$, corresponding to the
completely symmetric rank $m$ representation of $SU(3)$ that acts on
$x,y,$ and $z$ in the fundamental representation.

\paragraph{Single-Trace and Multi-Trace at $N\to \infty$: }
Having studied $f_\infty$, let us now look at the function $f_1$, 
generating the single-trace
operators for the case of one D3-brane on $\IC^3$. Clearly, the adjoint
fields $x,y,$ and $z$ are complex variables and not matrices and
therefore any product of two or more of these matrices is a
multi-trace operator. As a result, there are only 4 single-trace operators
in this case: the identity operator, $x,y,$ and $z$. We can therefore
use their representation in terms of $t_i, i=1,2,3$, sum them and
write:
\beq\label{f1C3}
f_1 (t_1,t_2,t_3) = 1+t_1+t_2+t_3.
\eeq

Next, we notice an interesting relation between $g_1$ and
$f_\infty$. Let us look at the set of operators of the form
$\tr(x^iy^jz^k)$ for the case in which the number of D3-branes is
$N \to \infty$. Each such operator is represented by the monomial $t_1^i
t_2^j t_3^k$ and can be thought of as a multi-trace operator for the
case of the number of D3-branes being $N=1$. This implies that $g_1$, the
generating function for multi-trace operators for one D3-brane is
equal to $f_\infty$, the generating function for single-trace
operators for infinitely many D3 branes,
\beq\label{g1finf}
g_1 = f_\infty \ .
\eeq

Can we now find some functional dependence between
$f_1$ and $g_1$? Combining expressions \eref{g1finf}, \eref{f1C3} and
\eref{C3}, we have
\bea
g_1(t_1,t_2,t_3) &=& \frac{1}{(1-t_1)(1-t_2)(1-t_3)} = \exp [ - \log
  (1-t_1) - \log (1-t_2) - \log (1-t_3) ] \cr &=& \exp \biggr(
\sum_{r=1}^\infty \frac{t_1^{r} + t_2^{r} + t_3^{r}}{r}\biggr ) = \exp
\biggr( \sum_{r=1}^\infty \frac{f_1(t_1^r,t_2^r,t_3^r)-1}{r}\biggr ).
\eea
The last relation
\beq\label{g1-f1}
g_1(t_1,t_2,t_3) = \exp \biggr( \sum_{r=1}^\infty
\frac{f_1(t_1^r,t_2^r,t_3^r)-1}{r}\biggr ) = f_\infty(t_1,t_2,t_3)
\eeq
turns out to be a key relation and is satisfied for any CY manifold,
toric or otherwise. The function
$g_1$ is then said to be the {\bf Plethystic Exponential}
of $f_1$. This relation in fact generalizes to any $N$ and we find
that $g_N$ is the plethystic exponential of $f_N$. We will discuss
this extensively in \sref{s:multi} and \sref{s:finiteN}.

We are now ready to write down the expression for the generating
function $g_\infty$ of multi-trace BPS GIO's in the chiral ring in
the $\cN=4$ theory, corresponding to $N \to \infty$ D3-branes on
$\IC^3$. It is again, the plethystic exponential, this time of
$f_\infty$ in \eref{C3}: 
\beq
g_\infty(t_1,t_2,t_3) = \exp \biggr( \sum_{r=1}^\infty
\frac{f_\infty(t_1^r,t_2^r,t_3^r)-1}{r}\biggr ) = \exp \biggr(
\sum_{r=1}^\infty \frac{
  \frac{1}{(1-t_1^r)(1-t_2^r)(1-t_3^r)}-1}{r}\biggr ) \ .
\eeq

Note that $g_\infty$ has an expansion
\beq
g_\infty(t_1,t_2,t_3) = \sum_{ijk} d_{ijk} t_1^i t_2^j t_3^k,
\eeq
where the coefficients $d_{ijk}$ are non-zero precisely when the
coefficients $c_{ijk}$ of $f_\infty$ are non-zero. However, while
$c_{ijk}$ can be at most $1$, $d_{ijk}$ has a very fast growth and in
fact grows exponentially. It is therefore a problem of interest to
find what is the large charge behavior of $d_{ijk}$.
We see that the multiplicity of BPS states for fixed $R$
charge, $R=\frac{2}{3}k$, is
\beq\label{g-C3}
g_\infty(t,t,t) = \exp \biggr( \sum_{r=1}^\infty \frac{
  \frac{1}{(1-t^r)^3}-1}{r}\biggr ) = \sum_{k=0}^\infty d_{k} t^k \ .
\eeq
We will present in \sref{s:asym} detailed discussions of how to obtain
$d_k$ for large $k$.

\paragraph{Single-Trace and Multi-Trace at Finite $N$: }
For finite $N$, the situation is in general much more involved. 
\comment{
Here,
however, the simplicity of $\IC^3$ allows us to write the generating
functions explicitly.
In fact, the function
$f_N$ is just the level-$N$ truncation of \eref{C3-f-ijk}:
\[
f_N(t_1,t_2,t_3) = \sum_{i,j,k=0}^{N} t_1^i t_2^j t_3^k \ . 
\]
}
Nevertheless, the multi-trace result $g_N$ can be obtained from $f_N$ by
plethystics. We will present the systematic treatment for
arbitrary singularities in \sref{s:finiteN}.
\section{Counting Gauge Invariants: Poincar\'e Series and Single-Trace}
\label{s:single}
\setall
Having stated our problem and enticed the reader with the example of
$\IC^3$, we are now ready to attack the general CY singularity. Our
strategy will be to first examine the simpler case of $N \to \infty$
and then the more involved case of finite $N$. 

Beginning with the large $N$ situation, we first
find the generating function $f$ for the
single-trace GIO's. Then, in \sref{s:multi},
we will show how the
  plethystic exponential (PE), extracts $g$, the generating function
for the multi-trace GIO's, from $f$. Indeed, because the multi-trace
GIO's are composed of products of the single-trace ones, PE
is expected to be a version of counting integer-partitions.
We would like to emphasize that the counting automatically encodes more
than merely the matter content, but, furtively, the superpotential as
well. In other words, we will be concerned with a true counting of the
GIO's with the F-term constraints automatically built in. We will
check in all examples below that this is indeed so by showing that the
moduli space is explicitly the CY 3-fold, as is required in D-brane
probe theories.

How, then, do we compute $f$ given the geometrical data of the CY? It
turns out that we could appeal to some known methods in
mathematics. In projective algebraic geometry, an important problem is
to count the number of generators of graded pieces of polynomial
rings, the generating functions of this type are called
{\bf Hilbert-Poincar\'e series}.

We shall borrow this terminology and refer to the function
$f$ for the single-trace GIO's as the Poincar\'e series for the
associated $\cN=1$ gauge theory; it shall soon be seen that this
appropriation is a conducive one.
In this section, we proceed stepwise along the various known classes of CY
singularities which the D3-brane can probe. We start with orbifolds
and see the Poincar\'e series in the mathematical sense is precisely
what is needed. Next, we address toric CY singularities; here, using
the techniques of $(p,q)$-webs and 2-dimensional tilings (dimers), we
construct $f$ from the toric diagram. Then, we proceed to the del
Pezzo family of singularities.

\subsection{Orbifolds and Molien Series}
Given a finite group, it is a classical problem to find the
generators of the ring of polynomial invariants under the group
action. The theory matured under E.~N\"other and T.~Molien
(cf.~e.g.~\cite{Yau}).
In our quiver gauge theory, the single-trace GIO's are polynomial
combinations of fields which are invariant under the group action.
Because we are assuming large $N$, no extra relations arise beside
these from the F-terms, and the problem of computing $f$ reduces to
simply counting the number of algebraically independent polynomials
one could construct of degree $n$ that are invariant under the
group.  The problem is a mathematical one and was solved by Molien;
the Poincar\'e series is named {\bf Molien series} in his honour.

Let us be concrete and specialise to the orbifolds of our concern,
viz., 3-dimensional CY orbifolds $\IC^3 / G$, with $G$ a discrete
finite subgroup of $SU(3)$. Such singularities were first classified
by \cite{Blichfeldt} and the D-brane quiver theories, constructed in
\cite{Hanany:1998sd}.
Let $G$ act on the coordinates $(x,y,z)$ of $\IC^3$.
Then, the question is: how many
algebraically\footnote{In fact linearly independent, 
  because any polynomial relation
  would change the total degree. Finding the polynomial relations is a
important one and will be subsequently addressed.}
independent polynomials are there of total degree $n$ in
$(x,y,z)$. The Molien series is given by
\beq\label{molien}
M(t;G) = \frac{1}{|G|}\sum_{g \in G} \frac{1}{\det(\II - t g)}
= \sum_{i=0}^\infty b_i t^i \ ,
\eeq
where the determinant is taken over the $3\times 3$ matrix
representation of the group elements. Upon series expansion, the
coefficients $b_i$ give the number of independent polynomials in degree
$i$. Hence, the $f$ we seek is simply $M(t;G)$.

We can remark one thing immediately. In \eref{molien} there is
only one variable $t$ instead of $(x,y,z)$ in our example $\IC^3$. 
The reason is
that for orbifold theories which descend from the ${\cal N}=4$ parent
every elementary field has R-charge ${2/3}$. The replacement $x,y,z\to
t$ tells us that \eref{molien} counts the single-trace GIO for
given R-charge.
Indeed,
as a first check, take $G = \II$, the trivial group. We immediately
find that
\beq\label{C3orb}
M(t; \II) = \frac{1}{\det(\II - t \II)} = \frac{1}{(1-t)^3} =
1 + 3\,t + 6\,t^2 + 10\,t^3 + 15\,t^4 + 21\,t^5 + \cO(t^6)
\ ,
\eeq
which agrees with \eref{C3} for the $\IC^3$ theory if one
set $t_i = t$. Thus, the Molien series counts invariants of {\it
total} degree in $x,y,z$ whereas \eref{C3} counts the degree of the
three variables individually. In the next subsection, we shall refine
the Molien series by straight-forwardly generalising the dummy variable
$t$ to a triple $(t_{1,2,3})$.

Emboldened by this check, let us go on to a non-trivial example,
the binary dihedral group $\hat{D}_4$ of 8
elements. This is a subgroup of $SU(2) \subset SU(3)$ and is a member of
the ADE-series of CY two-fold (K3) singularities (cf.~\cite{slodowy}).
We can think of this as a $\IC^3$ orbifold with one coordinate, say
$z$, held fixed. The gauge theory is the well-known $\cN=2$ D-type
quiver
(q.v.~\cite{Johnson:1996py,Kachru:1998ys,Lawrence:1998ja,Hanany:1998sd}).

This group is generated (we use the standard notation that
$\gen{x_1,\ldots,x_n}$ is the finite group generated by the
list of matrices $x_i$) as
\beq
\hat{D}_4 = \gen{\mat{-i & 0 \\ 0 & i}, \mat{0 & i \\ i & 0}} \ ,
\eeq
acting on $(x,y) \in \IC^2$. We can readily compute the Molien series to be
\bea\label{mol-D4}
M(t,\hat{D}_4) &=& \frac18(\frac{6}{1 + t^2} + \frac{1}{1 - 2\,t + t^2} +
  \frac{1}{1 + 2\,t + t^2}) \\ \nn
&=& 1 + 2\,t^4 + t^6 + 3\,t^8 + 2\,t^{10} + 4\,t^{12} + 3\,t^{14} +
  5\,t^{16} + 4\,t^{18} +  6\,t^{20} + O(t^{22}) \ .
\eea
This dictates that there are two invariants at degree 4, one at degree
6, etc.

Now, one can actually determine the invariants explicitly, which gives
us another check. First, an important theorem due to N\"other states
that (cf.~e.g.~\cite{Yau}):
\begin{theorem}\label{noether}
The polynomial ring of invariants is finitely generated and the degree
of the generators is bounded by $|G|$.
\end{theorem}
Therefore, though the Molien series is infinite, with increasingly
more invariants arising at successive degree with them
being linearly independent at each total degree, there will
be highly non-trivial algebraic relations amongst the ones at
different degree. The power of Theorem \ref{noether} is that one
needs to find invariants at most up to degree equal to the order of
the group; all higher degree invariants are polynomials in these basic
ones.

Hence, we only need to find a finite number of invariants, which can
be determined explicitly due to an averaging technique of
O.~Reynolds (cf.~e.g.~\cite{Yau}). 
Given any polynomial $F(x)$, one can define
the so-called {\bf Reynolds operator}
\beq\label{reynolds}
R_G[F(x)] := \frac{1}{|G|} \sum_{g \in G} F(g \circ x) \ .
\eeq
Then, the polynomial $R_G[F(x)]$ is invariant by construction. We can
then list all monomials of a given degree, apply \eref{reynolds} to
each and obtain the invariants at the said degree; the number thereof
should agree with what \eref{molien} predicts.

Applying the above discussion to our example of $\hat{D}_4$, we obtain
the following invariant polynomials for the first few degrees:
\beq\label{D4inv}\ba{c|c}
\mbox{degree} & \mbox{invariant polynomials} \\ \hline
4 & x^2y^2, \quad \frac12(x^4 + y^4) \\
6 & \frac12xy(x^4 - y^4) \\
8 & x^4y^4, \quad \frac12x^2y^2(x^4 + y^4), \quad \frac12(x^8 + y^8)
\ea\eeq
We remark that there are no invariants of lower degree (except
trivially the identity) and that the
number of independent
invariants indeed agree with the series expansion of \eref{mol-D4}.
Immediately, one sees some trivial relations such as $x^4y^4 =
(x^2y^2)^2$. Using Gr\"obner basis algorithms \cite{m2}, one can show
that the above ring of 6 invariants can be further reduced to 3. In
other words, the ring of invariant polynomials, $\IC[x,y]^{\hat{D}_4}$,
is generated by 3 so-called {\bf primitive} ones:
\beq
v = \frac12(x^4 + y^4), \quad
w = x^2y^2, \quad
u = \frac12xy(x^4 - y^4) \ .
\eeq

Finding relations among these polynomials is known as the {\bf syzygy
  problem} and is, again, a classical problem dating to at least
Hilbert. The modern solution is, as above, to use Gr\"obner bases. The
reader is referred to \cite{grob} for a pedagogical application of
syzygies and Gr\"obner basis to $\cN=1$ gauge theories and to
  \cite{compAG} 
in the context of moduli stabilisation. We will return to syzygies
later in the paper. For the present example, we find the relation
\beq\label{eqD4}
v^2w - w^3 = u^2 \ .
\eeq
This is a comforting result. Indeed, invariant theory tells us that
\qq{The defining equation of an orbifold is the syzygy of the
  primitive invariants.}
We recognise \eref{eqD4} as precisely the defining equation
  \cite{slodowy} for the affine variety $\IC^2/\hat{D}_4$.

\subsubsection{ADE-Series}\label{s:ADE}
For completeness, let us compute (making extensive use of \cite{m2,gap}) 
the Molien series for the discrete
subgroups of $SU(2)$. We find that
\beq\ba{|c|c|c|c|c|} \hline
G \subset SU(2) & |G| & \mbox{Generators} & \mbox{Equation}
& \mbox{Molien }
M(t; G) \\ \hline
\hat{A}_{n-1} & n & \gen{\mat{\omega_n & 0 \\ 0 & \omega_{n}^{-1}}} &
uv = w^n
&
{\Large \mbox{$\frac{(1+t^n)}{(1-t^2)(1-t^n)}$}}
\\ \hline
\hat{D}_{n+2} & 4n & \gen{\mat{\omega_{2n} & 0 \\ 0 & \omega_{2n}^{-1}},
  \mat{0 & i \\ i & 0}} &
% D-series = A-Seres + n / (1+t^2) since the off-diagonal elements
% are {{0, i w_n^k}, {i w_n^{-k}, 0}} are Det(1-t x) is simple
% Also, the cyclic group generator must be even-order
u^2 + v^2w = w^{n+1}
&
{\Large \mbox{$\frac{(1+t^{2n+2})}{(1-t^4)(1-t^{2n})}$}}
\\ \hline
\hat{E}_6 & 24 & \gen{S,T} & u^2+v^3+w^4=0
&
{\Large \mbox{$\frac{1 - t^4 + t^8}{1 - t^4 - t^6 +  t^{10}}$}}
\\ \hline
\hat{E}_7 & 48 & \gen{S,U} & u^2+v^3+vw^3=0 &
\ba{c}\\
{\Large \mbox{$\frac{1 - t^6 + t^{12}}{1 - t^6 - t^8 + t^{14}}$}}\ea
\\ \hline
\hat{E}_8 & 120 & \gen{S,T,V} & u^2+v^3+w^5=0 &
\ba{c}\\
{\Large \mbox{$\frac{1 + t^2 - t^6 - t^8 - t^{10} + t^{14} + t^{16}}
{1 + t^2 - t^6 - t^8 - t^{10} - t^{12} + t^{16} + t^{18}}$}}
\ea
\\ \hline
\ea\eeq
where we have defined $\omega_n := e^{\frac{2\pi i}{n}}$ and
\beq
\ba{l}
S := \frac12\mat{-1+i & -1+i \\ 1+i & -1-i}, \quad
T := \mat{i & 0 \\0& -i}, \\
U := \frac{1}{\sqrt{2}}\mat{1+i & 0 \\ 0 & 1-i}, \quad
V := \mat{\frac{i}{2} & \frac{1-\sqrt{5}}{4}- i \frac{1+\sqrt{5}}{4} \\
  -\frac{1-\sqrt{5}}{4}-i \frac{1+\sqrt{5}}{4} & -\frac{i}{2}} \ .
\ea
\eeq
We have also used the identity
\bea \nn
\comment{
\sum_{k=0}^{n-1} \frac{1}{1 - t \omega_n^k}
&=& \sum_{k=0}^{n-1} \sum_{j=0}^\infty t^j \omega_n^{jk}
= \sum_{j=0}^\infty t^j \frac{1 - \omega_n^{jn}}{1-\omega_n^j}
= \sum_{j=0}^\infty t^j n \delta_{j, n\IZ_{\ge 0}}
= \frac{n}{1-t^n} \\ \nn
}
\sum_{k=0}^{n-1} \frac{1}{(1 - t \omega_n^k)(1 - t \omega_n^{-k})}
&=&
\sum_{j=0}^\infty \sum_{m=0}^\infty t^{j+m} n \delta_{j, n\IZ}
= n \sum_{m=0}^\infty
\left(
\sum_{\beta=0}^\infty t^{2m+n\beta} + \sum_{\beta=1}^\infty
t^{2m-n\beta}
\right) \\ \nn
&=& \frac{n}{1-t^2}\left(
\frac{1}{1-t^n} + \frac{1}{t^{-n}-1}
\right) \ . \\
\eea

\subsubsection{Valentiner: A Non-Abelian $SU(3)$
  Example}\label{s:delta27}
Having warmed up with the 2-dimensional CY orbifolds, we are ready to
study the proper subgroups of $SU(3)$ \cite{Hanany:1998sd}.
The simplest, most well-known, non-trivial, non-Abelian discrete
subgroup of $SU(3)$ is perhaps the Valentiner group, otherwise known
as $\Delta(3\cdot 3^2)$ (or, sometimes known as the Heisenberg group
for 3 elements, as recently studied in \cite{Burrington:2006uu}), 
defined as
\beq
\Delta(27) :=
\gen{
\mat{\omega_3 & 0 & 0 \\ 0 & 1 & 0 \\ 0 & 0 & \omega_3^{-1}},
\mat{1 & 0 & 0 \\ 0 & \omega_3 & 0 \\ 0 & 0 & \omega_3^{-1}},
\mat{0&1&0 \\  0&0&1 \\ 1&0&0}
} \ .
\eeq
Let us investigate this group in some detail; we shall return to this
group later in the paper.
The Molien series is readily computed to be
\beq\label{Mdel27}
M(t; \Delta(27)) =
\frac{-1 + t^3 - t^6}{{\left( -1 + t^3 \right) }^3}
= 1 + 2\,t^3 + 4\,t^6 + 7\,t^9 + 11\,t^{12} + 16\,t^{15} + 22\,t^{18}
+ \ldots \ .
\eeq

To find the defining equation (syzygies),
Theorem \ref{noether} tells us that we need
only go up to degree 27 here, a total of 174 invariants, of degrees
$0,3,6,\ldots, 24, 27$. Using Gr\"obner techniques \cite{m2},
we find that there are only 4 nontrivial
generators for these 174 polynomials
(we have scaled the expressions by an over-all 3):
\beq\label{invdelta27}
\{
m = 3xyz, \quad
n = x^3+y^3+z^3, \quad
p = x^6 + y^6 + z^6, \quad
q = x^3y^6 + x^6z^3 + y^3z^6
\} \ .
\eeq
We then find a single relation in $\IC[m,n,p,q]$:
\beq\label{eqdelta27}
8\,m^6 + m^3\,\left( -48\,n^3 + 72\,n\,p + 72\,q \right)
      + 81\,\left( {\left( n^2 - p \right) }^3 -
     4\,n\,\left( n^2 - p \right) \,q + 8\,q^2 \right)
= 0 \ .
\eeq
Therefore, $\IC^3/\Delta(27)$ is a complete intersection, given by a single
(Calabi-Yau) hypersurface in $\IC^4$.

%========================================
%\subsubsection{All SU(3) Groups}
% Should be able to find these in the book.

%
% TORIC
%
\subsection{Toric Varieties}\label{s:toric}
Having studied the first class of CY singularities, viz., the
orbifolds, in some detail, let us move onto the next, and recently
much-revived, class of geometries, the toric singularities.
It turns out that here one can
also write the Poincar\'e series $f$ explicitly, now in terms of the
combinatorics of the given toric diagram $D$ \cite{Martelli:2006yb}. 
Mathematically, this is
a nice extension of the Molien series.

We first draw the graph dual\footnote{
  Incidentally, we remark that a convenient way of finding the dual 
  $(p,q)$-web given a toric diagram is to take, for each pair of
  toric points, their cross-product, which then gives a vector
  perpendicular to the plane defined by the two said points.
} of the triangulation of $D$; 
this is the $(p,q)$-web \cite{Aharony:1997bh}, a
skeleton of tri-valent vertices indexed by $i \in V$. At each vertex $i$,
the $j$-th (for $j=1,2,3$) of the three coincident edges
has charge $\vec{a}_{ij}$ with $\vec{a}$ a three-vector indexed by
$k$, signifying the 3 charges. We remark
that toric Calabi-Yau threefolds have three-dimensional toric diagrams
whose endpoints are co-planar and this is why $D$ and the dual
$(p,q)$-web are usually drawn on the plane. Here, we need to restore
the full coordinates of the 3-dimensional 
toric diagram; whence, $\vec{a}$ has 3
components. With this notation, the Poincar\'e series for $D$ is
(cf.~Eq.~(7.24-5) of \cite{Martelli:2006yb} and also \cite{brion} for 
interesting mathematical perspectives):
\beq\label{p-toric}
P(t_1, t_2, t_3; D) = \sum\limits_{i \in V}
\prod\limits_{j=1}^3
\frac{1}{1 - t_1^{a_{ij}^1}t_2^{a_{ij}^2}t_3^{a_{ij}^3}} \ .
\eeq

Before we proceed, let us remark on the charges of coordinates
$t_1,t_2,t_3$. For toric varieties, we have
three $U(1)$ global symmetries: one is R-charge and the other two,
flavor charges. In general, each coordinate $t_i$ is charged under
all three $U(1)$. For example, the R-charge of $t_2$ is given by
the inner product of $(0,1,0)$ and the {\bf Reeb Vector}
$V_R=(b_1,b_2,b_3)$. We recall that in the AdS/CFT correspondence 
the $U(1)$ R-symmetry is dual to a special Killing vector, the
so-called  Reeb vector (cf.~e.g.~\cite{Martelli:2005tp}), 
which can be expanded as $V_R= \sum\limits_{i=1}^3
b_i {\partial \over \partial \phi_i}$, where $\phi_i$ are the
coordinates parametrising the $T^3$-toric action. It is a very
important quantity in toric geometry.

It is possible to make coordinate
transformation $(t_1,t_2,t_3)\to (\W t_1, \W t_2, \W t_3)$ such that
each coordinate $\W t_i$ is charged under one and only one $U(1)$.
However, the charges of these new coordinates $\W t_i$ in general are
not even rational numbers (for example the R-charge of $dP_2$), so
it is not proper to use this new $\W t_i$ coordinate to do the
Poincar\'e series expansion, which must have integer powers.
Furthermore, as we have seen in the example of $\IC^3$, sometimes we
 want to find the generating function of only one $U(1)$ charge,
 for example, the R-charge. To do so, we merely make the substitution
 $(t_1,t_2,t_3) \to (t^a, t^b, t^c)$ for given $a,b,c \in \IZ_{\ge
   0}$ and the expression will be simplified
 considerably. In a lot of cases, the interesting $U(1)$ is a linear
 combination of all three $U(1)$'s as we will see shortly.

Returning to \eref{p-toric},
we have some immediate checks. First, we recall that
all Abelian orbifolds of $\IC^3$ (including $\IC^3$ itself) are toric.
For example, the toric diagram for $\IC^3$ is a triangle with vertices
$(1,0,0)$, $(0,1,0)$ and $(0,0,1)$. The dual graph, i.e., the
$(p,q)$-web, has a single
vertex, with three edges in the directions $(1,0,0)$, $(0,1,0)$ and
$(0,0,1)$ respectively:
\[\epsfxsize=5in\epsfbox{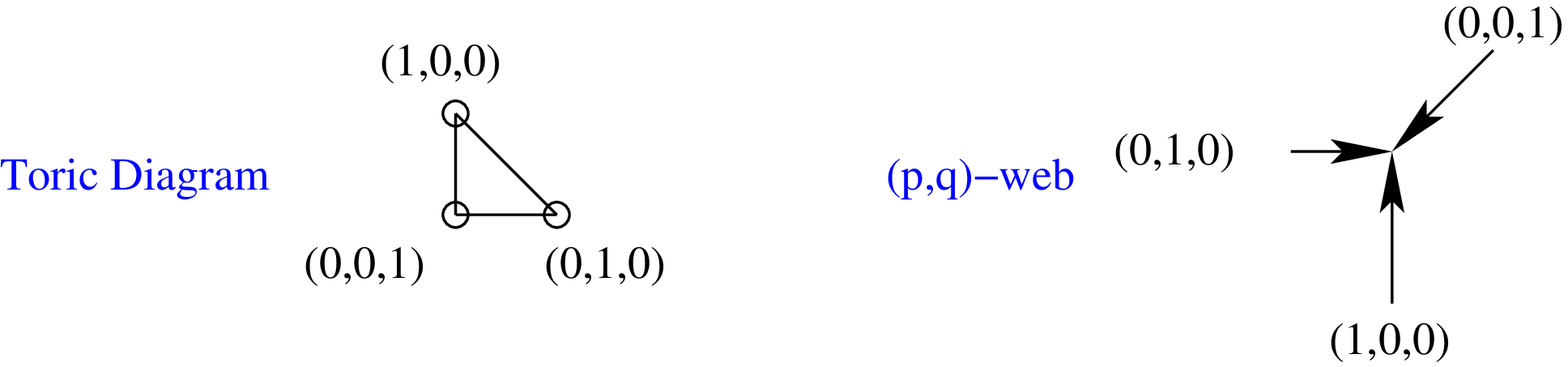}\]
Hence,
\bea\nn
P(t_1, t_2, t_3; \IC^3) &=&
\frac{1}{1-t_1^1 t_2^0 t_3^0}\frac{1}{1-t_1^0 t_2^1
  t_3^0}\frac{1}{1-t_1^0 t_2^0 t_3^1} \\
&=& \frac{1}{(1-t_1)(1-t_2)(1-t_3)} =
\sum_{i,j,k} t_1^i t_2^j t_3^k \ .
\label{sGIOC3}
\eea
This is precisely the result \eref{C3} obtained from conventional
methods in \sref{s:c3eg}.

\begin{figure}[h]
\epsfxsize=6.5in\epsfbox{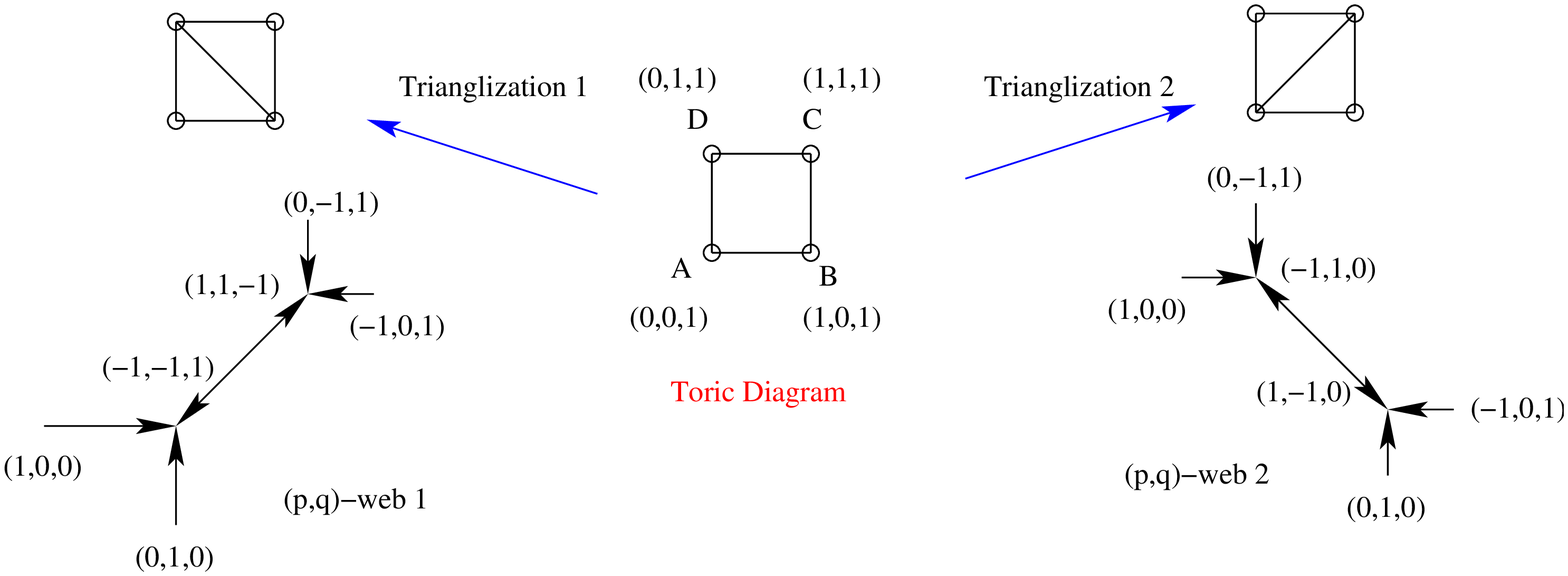}
\caption{{\sf The toric data for the conifold $\cC$. There are two
  triangulations, related to by flops, and thus two $(p,q)$-webs. We
  see in the text that they lead to the same counting.}}
\label{f:coniflop}
\end{figure}

For a less trivial example, take the conifold ${\cal C}$
(cf.~\cite{Martelli:2006yb} as well as an earlier result in
\cite{Nekrasov:2004vw}).
The toric diagram has 4 points, with coordinates
\beq
A=(0,0,1),\qquad B=(1,0,1), \qquad C=(1,1,1), \qquad D=(0,1,1) \ ,
\eeq
as shown in the center of \fref{f:coniflop}.
There are two triangulations, giving two $(p,q)$-webs upon
dualising; the two are related by flop transitions.
Of course, we need to prove the counting is independent of such choices. 
Indeed, as the conifold
is the building block to all flops in toric varieties, if we show that
$f$ is the same for the two $(p,q)$-webs, this would be true for all
toric diagrams and thus we would be at liberty to make any choice of
$(p,q)$-web.
First, take the left one, given by the two triangles $ABD$ and
$BCD$. This gives us 2 vertices, with $(p,q)$-charges of, respectively,
$\{(0,1,0),~(1,0,0),~(-1,-1,1)\}$ and
$\{(0,-1,1),~(-1,0,1),~(1,1,-1)\}$. Thus \eref{p-toric} gives us
\beq\label{coni1}
P(x,y,z;~~\cC) = 
\frac{1}{\left( 1 - \frac{x\,y}{z} \right) \,
     \left( 1 - \frac{z}{x} \right) \,
     \left( 1 - \frac{z}{y} \right) } + 
  \frac{1}{\left( 1 - x \right) \,
     \left( 1 - y \right) \,
     \left( 1 - \frac{z}{x\,y} \right) } \ . 
\eeq
The second
trianglization is given by $ACD$ and $ABC$, giving us
\beq\label{coni2}
\frac{1}{\left( 1 - \frac{x}{y} \right) \,
     \left( 1 - y \right) \,
     \left( 1 - \frac{z}{x} \right) } + 
  \frac{1}{\left( 1 - x \right) \,
     \left( 1 - \frac{y}{x} \right) \,
     \left( 1 - \frac{z}{y} \right) } \ .
\eeq 
It is easy
to see that the two expressions \eref{coni1} and \eref{coni2}
are the same. Thus indeed the
generating function is independent of how we triangulate and how the
dual $(p,q)$-web is obtained \cite{Martelli:2006yb}.

%
%===================
\subsubsection{Refinement: $U(1)$-charges and Multi-degrees}\label{s:refine}
We see from \eref{p-toric} that for toric varieties the counting is
more refined than the Molien series \eref{molien} as the latter only
counts invariants of total-degree. There seems to be a
straight-forward generalisation.
In order to get the number of single-trace GIO's given the R-charge
of each field, the Molien counting seems to be refinable to counting
the number of independent polynomials of a given multi-degree $(i_1,
\ldots, i_3)$. This is done by generalising the
Molien series to:
\beq
M(t,G) = \frac{1}{|G|}\sum_{g \in G} \frac{1}{\det(\II -
  \mbox{diag}(t_1, \ldots, t_k) \cdot g)}
= \sum_{i_1,\ldots, i_k}
  b_{i_1, \ldots, i_k} t_1^{i_1}\ldots t_k^{i_k} \ .
\eeq

The caveat is that
now the coefficients $b_{i_1, \ldots, i_k}$ are no
longer guaranteed to be integers for general groups.
This corresponds to the fact that
the invariants are not monomial in general, but, rather,
polynomial. For example, in \eref{D4inv}, at degree 6, there is a
single invariant, built of the sum of two monomials, of multi-degree
$(5,1)$ and $(1,5)$, each of which is {\it not} an invariant.

Therefore, this refinement only makes sense
in case there is a corresponding
conserved charge associated with a $U(1)$ that is part of the isometry
of the CY manifold.
The isometry of $\IC^3$ is $SU(4)$ with rank 3. There is thus a
maximum
of 3 charges corresponding to the maximal subgroup of the  isometry
group of the CY manifold.
If the manifold is toric then there is a $T^3$ fibration and therefore  a
total of three $U(1)$ charges and the index would be a function of 3
variables.
If the manifold is not toric then in many cases the isometry group
has a rank smaller than 3 and in most cases in fact is absent.
Nevertheless there is at least one charge, counting the R charge,
that corresponds to the choice of complex structure of the manifold.

To summarize, there are some cases in which the rank of the isometry
group is 2 and in most cases the rank is 1. All cases in which the
rank is less than 3 are non-toric.
An example for a manifold with rank 2 is the set of complete
intersection manifolds of the form
$x^2+y^2+z^2+w^k = 0$.
Is has a clear $SO(3)$ isometry acting on the first 3 coordinates and
together with the natural degree of the variables form the isometry
group $SU(2) \times U(1)$.
For the case $k=2$ the isometry grows to $SO(4) \times U(1)$
and having rank 3 indeed confirms that the manifold is toric - the
familiar conifold. We will study this geometry again in
\sref{s:xyz-w}.

An example of a manifold of rank 1 is any $\IC^3$-orbifold 
with a full non-abelian
subgroup $\Gamma$ of $SU(3)$. For $\Gamma$ in $SU(2)$, 
we still have $\cN=2$ SUSY, so the global
isometry is $SU(2) \times U(1)$, of rank 2.
\comment{In order to see this, one should consider non-holomorphic
  operators.} 
Indeed though there is no refinement for 3 charges, there should still be a
refinement of two charges since the rank is 2.
The first charge will denote the Cartan charge of the $SU(2)$.
We could, for example, define a degree which counts how many $x$'s and
$y$'s together, and another degree which counts how many $z$'s.
This is the reason that in trying
to implement the refinement on $\hat{D}_4$ we found factors of
$1/2$. There is simply no corresponding conserved charge which
corresponds to this generalization.

\subsubsection{The $Y^{p,q}$ Family}
\begin{figure}
\epsfxsize=6in\epsfbox{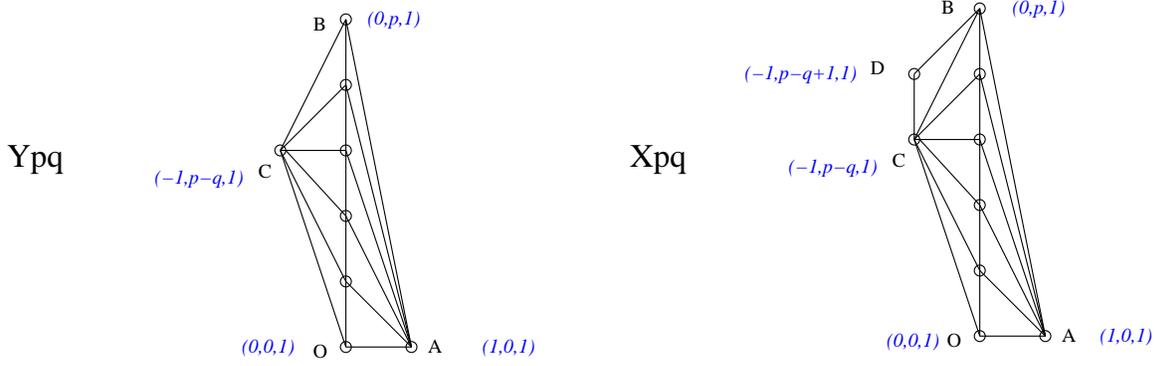}
\caption{{\sf The toric diagrams for the spaces $Y^{p,q}$ and
    $X^{p,q}$.}}
\label{f:XY}
\end{figure}

An infinite family of toric CY 3-folds which has recently attracted
much attention, because of the discovery of explicit CY metric
thereon, is the $Y^{p,q}$'s (cf.~e.g.,\cite{Martelli:2004wu,Benvenuti:2004dy,Benvenuti:2004wx}). 
Let us now do the counting for these.
The toric data is given by $O=(0,0,1)$, $A=(1,0,1)$, $B=(0,p,1)$ and
$C=(-1,p-q,1)$. We have drawn it at the left hand side of \fref{f:XY}.
As indicated, we take the trianglization as connecting the point
$T_a=(0,a,1)$ to $A$ and $C$ with $a=1,...,p$ (so $T_p=B$). Now we
have $2p$ triangles given by $T_a A T_{a+1}$ and $T_a C T_{a+1}$,
$a=0,...,p-1$. For triangle $T_a A T_{a+1}$ we have the following
charges and corresponding term \comment{where I have used $q=z^{-1}$
as in del Pezzo case}
\beq\label{Ypq-1}
\{(1,0,0), ~~(a,1,-a),~~~(-a-1,-1,a+1)\}\Rightarrow 
\frac{1}{\left( 1 - x \right) \,
    \left( 1 - \frac{x^a\,y}{z^a} \right) \,
    \left( 1 - \frac{x^{-1 - a}\,z^{1 + a}}{y} \right) } \ .
\eeq
For triangle $T_a C T_{a+1}$ we have
\beq\label{Ypq-2}
\scriptsize{\mbox{$\{(-1,0,0),~~(-a+(p-q),1,-a),~~((q-p)+a+1,-1,a+1)\}$}} 
\Rightarrow
\frac{1}{\left( 1 - \frac{1}{x} \right) \,
    \left( 1 - \frac{x^{-a + p - q}\,y}{z^a} \right) \,
    \left( 1 - \frac{x^{1 + a - p + q}\,z^{1 + a}}
       {y} \right) } \ .
\eeq
Putting these together we have
\beq\label{p-ypq}
\hspace{-1cm}
f(x,y,q;~Y^{p,q})
= \sum_{a=0}^{p-1}
\frac{1}{\left( 1 - x \right) \,
    \left( 1 - \frac{x^a\,y}{z^a} \right) \,
    \left( 1 - \frac{x^{-1 - a}\,z^{1 + a}}{y} \right) } +
\frac{1}{\left( 1 - \frac{1}{x} \right) \,
    \left( 1 - \frac{x^{-a + p - q}\,y}{z^a} \right) \,
    \left( 1 - \frac{x^{1 + a - p + q}\,z^{1 + a}}
       {y} \right) } \ .
\eeq
Knowing $Y^{p,q}$, it is easy to go on to $X^{p,q}$. It differs
therefrom by the addition of one point $(-1,p-q+1,1)$ in the toric
diagram. So we use the above trianglization of $Y^{p,q}$, plus one
more triangle given by $(0,p,1)$, $(-1,p-q+1,1)$ and $(-1,p-q,1)$.
This one gives the following vector and hence a new term to
\eref{p-ypq}:
\[
\{ (1,0,1), (q-1,-1,p), (-q,1,-p)\} \quad \Rightarrow \quad 
\frac{1}{\left( 1 - x\,z \right) \,
    \left( 1 - \frac{y}{x^q\,z^p} \right) \,
    \left( 1 - \frac{x^{-1 + q}\,z^p}{y} \right) } \ .
\]

%
%============= del Pezzo
%
\subsection{The Del Pezzo Family}
The last category of CY singularities widely studied in D-brane gauge
theories is the cone over the 9 del Pezzo surfaces. These surfaces are
$\IP^2$ blown up at $n$ generic points; the cone is CY if
$n=0,\ldots,8$. There is a close cousin to this family, viz, the
zeroth Hirzebruch surface $F_0$, which is simply $\IP^1 \times \IP^1$
and the cone over which is also CY.
It is well-known that for $dP_{n=0,1,2,3}$ and for $F_0$, the space is
actually toric (q.v.~\cite{Feng:2000mi}). 
Therefore, we can use \eref{p-toric} to obtain the
following:
\beq\label{f-dP0123}
\ba{rcl}
P(z,x,y; dP_0) &=&
\frac{1}{\left( 1 - x \right) \,\left( 1 - y \right) \,
     \left( 1 - \frac{z}{x\,y} \right) } + 
  \frac{1}{\left( 1 - \frac{1}{x} \right) \,
     \left( 1 - \frac{y}{x} \right) \,
     \left( 1 - \frac{x^2\,z}{y} \right) } + 
  \frac{1}{\left( 1 - \frac{1}{y} \right) \,
     \left( 1 - \frac{x}{y} \right) \,
     \left( 1 - \frac{y^2\,z}{x} \right) }\\
P(z,x,y; F_0) &=&
\frac{1}{\left( 1 - x \right) \,\left( 1 - y \right) \,
     \left( 1 - \frac{z}{x\,y} \right) } + 
  \frac{1}{\left( 1 - \frac{1}{x} \right) \,
     \left( 1 - y \right) \,
     \left( 1 - \frac{x\,z}{y} \right) } + 
  \frac{1}{\left( 1 - x \right) \,
     \left( 1 - \frac{1}{y} \right) \,
     \left( 1 - \frac{y\,z}{x} \right) } + 
  \frac{1}{\left( 1 - \frac{1}{x} \right) \,
     \left( 1 - \frac{1}{y} \right) \,
     \left( 1 - x\,y\,z \right) }\\
P(z,x,y; dP_1) &=&
\frac{1}{\left( 1 - x \right) \,
     \left( 1 - \frac{y}{x} \right) \,
     \left( 1 - \frac{z}{y} \right) } + 
  \frac{1}{\left( 1 - \frac{1}{x} \right) \,
     \left( 1 - y \right) \,
     \left( 1 - \frac{x\,z}{y} \right) } + 
  \frac{1}{\left( 1 - x \right) \,
     \left( 1 - \frac{x}{y} \right) \,
     \left( 1 - \frac{y\,z}{x^2} \right) } + 
  \frac{1}{\left( 1 - \frac{1}{x} \right) \,
     \left( 1 - \frac{1}{y} \right) \,
     \left( 1 - x\,y\,z \right) }\\
P(z,x,y; dP_2) &=&
\frac{1}{\left( 1 - \frac{x}{y} \right) \,
     \left( 1 - y \right) \,
     \left( 1 - \frac{z}{x} \right) } + 
  \frac{1}{\left( 1 - x \right) \,
     \left( 1 - \frac{y}{x} \right) \,
     \left( 1 - \frac{z}{y} \right) } + 
  \frac{1}{\left( 1 - \frac{1}{x} \right) \,
     \left( 1 - y \right) \,
     \left( 1 - \frac{x\,z}{y} \right) } + \\
&& + \frac{1}{\left( 1 - x \right) \,
     \left( 1 - \frac{1}{y} \right) \,
     \left( 1 - \frac{y\,z}{x} \right) } + 
  \frac{1}{\left( 1 - \frac{1}{x} \right) \,
     \left( 1 - \frac{1}{y} \right) \,
     \left( 1 - x\,y\,z \right) }\\
P(z,x,y; dP_3) &=&
\frac{1}{\left( 1 - \frac{1}{y} \right) \,
     \left( 1 - x\,y \right) \,
     \left( 1 - \frac{z}{x} \right) } + 
  \frac{1}{\left( 1 - \frac{1}{x\,y} \right) \,
     \left( 1 - y \right) \,\left( 1 - x\,z \right) } + 
  \frac{1}{\left( 1 - \frac{1}{x} \right) \,
     \left( 1 - x\,y \right) \,
     \left( 1 - \frac{z}{y} \right) } + \\
 && + \frac{1}{\left( 1 - x \right) \,
     \left( 1 - y \right) \,
     \left( 1 - \frac{z}{x\,y} \right) } + 
  \frac{1}{\left( 1 - x \right) \,
     \left( 1 - \frac{1}{x\,y} \right) \,
     \left( 1 - y\,z \right) } + 
  \frac{1}{\left( 1 - \frac{1}{x} \right) \,
     \left( 1 - \frac{1}{y} \right) \,
     \left( 1 - x\,y\,z \right) } \ .
\ea\eeq 
%In these expressions $q={1\over z}$ counts the number of
%states along a plane which is perpendicular to the internal point in
%the toric diagram; $x$ and $y$ are the coordinates along this plane.
We include toric diagrams and the dual $(p,q)$-webs here for
reference:\\
\[\ba{ccc} 
\ba{l}\epsfxsize=3in\epsfbox{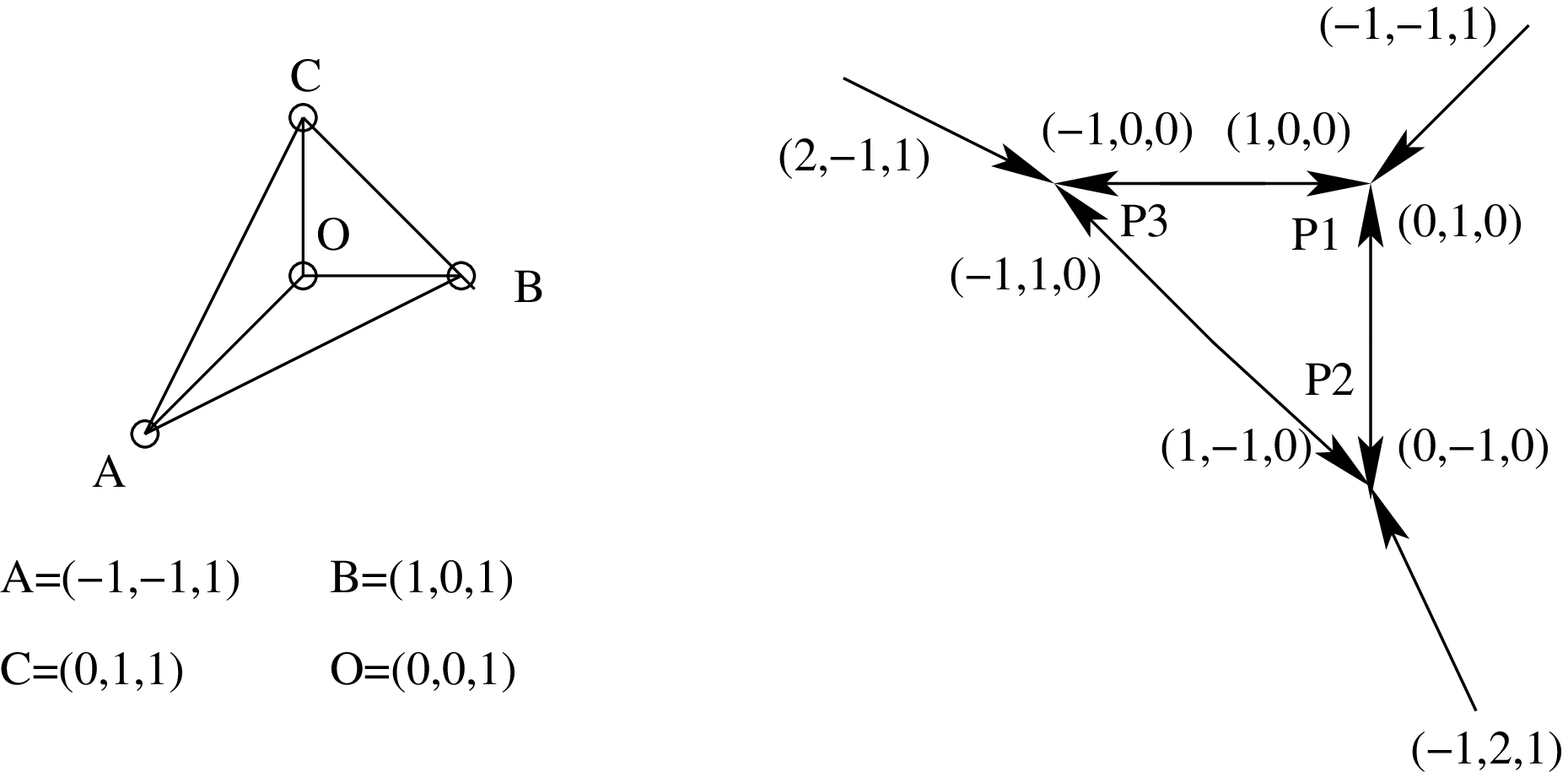}\ea &&
\ba{l}\epsfxsize=3in\epsfbox{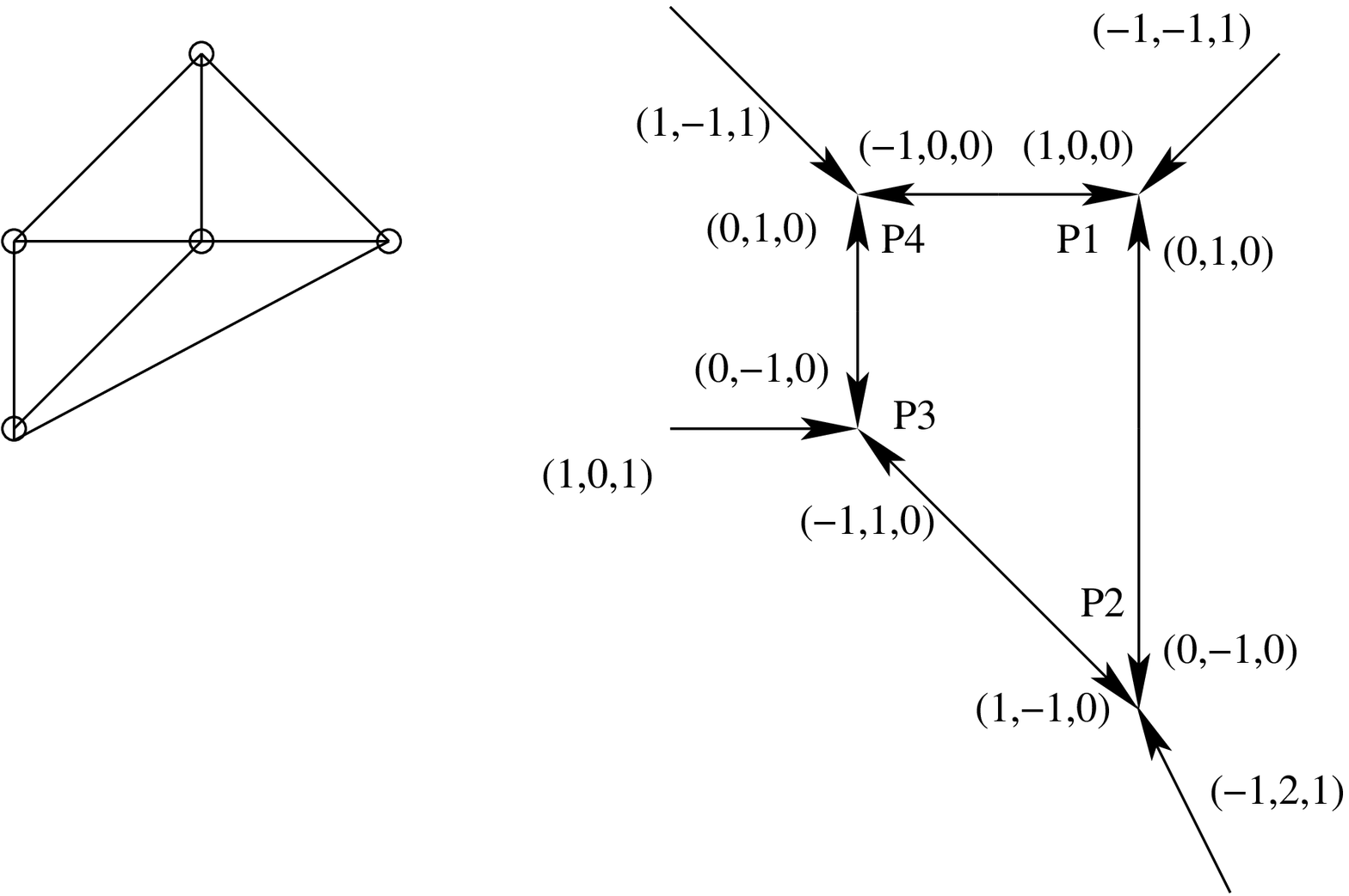}\ea \\
(a)~~dP_0 && (b)~~dP_1\\
\ba{l}\epsfxsize=3in\epsfbox{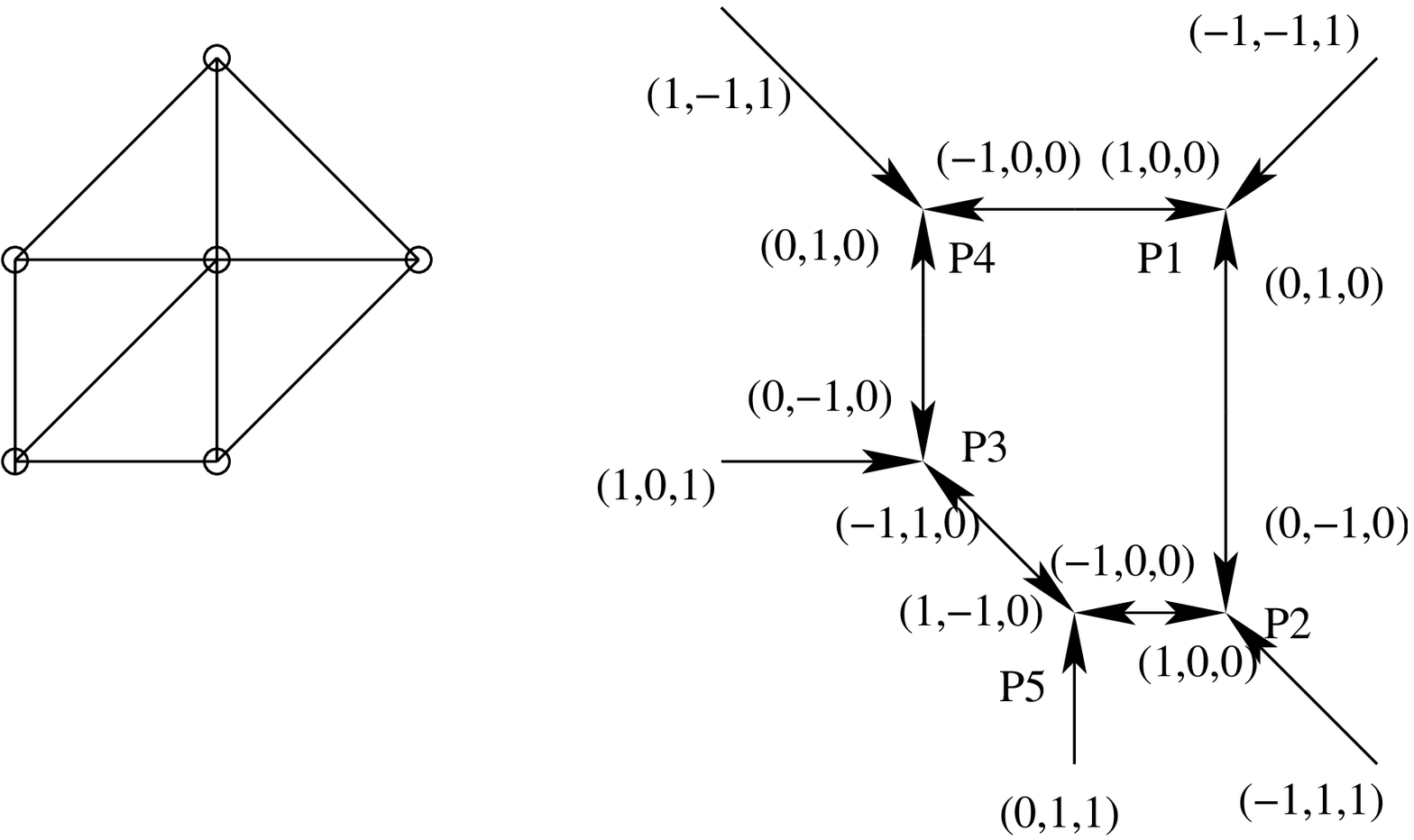}\ea &&
\ba{l}\epsfxsize=3in\epsfbox{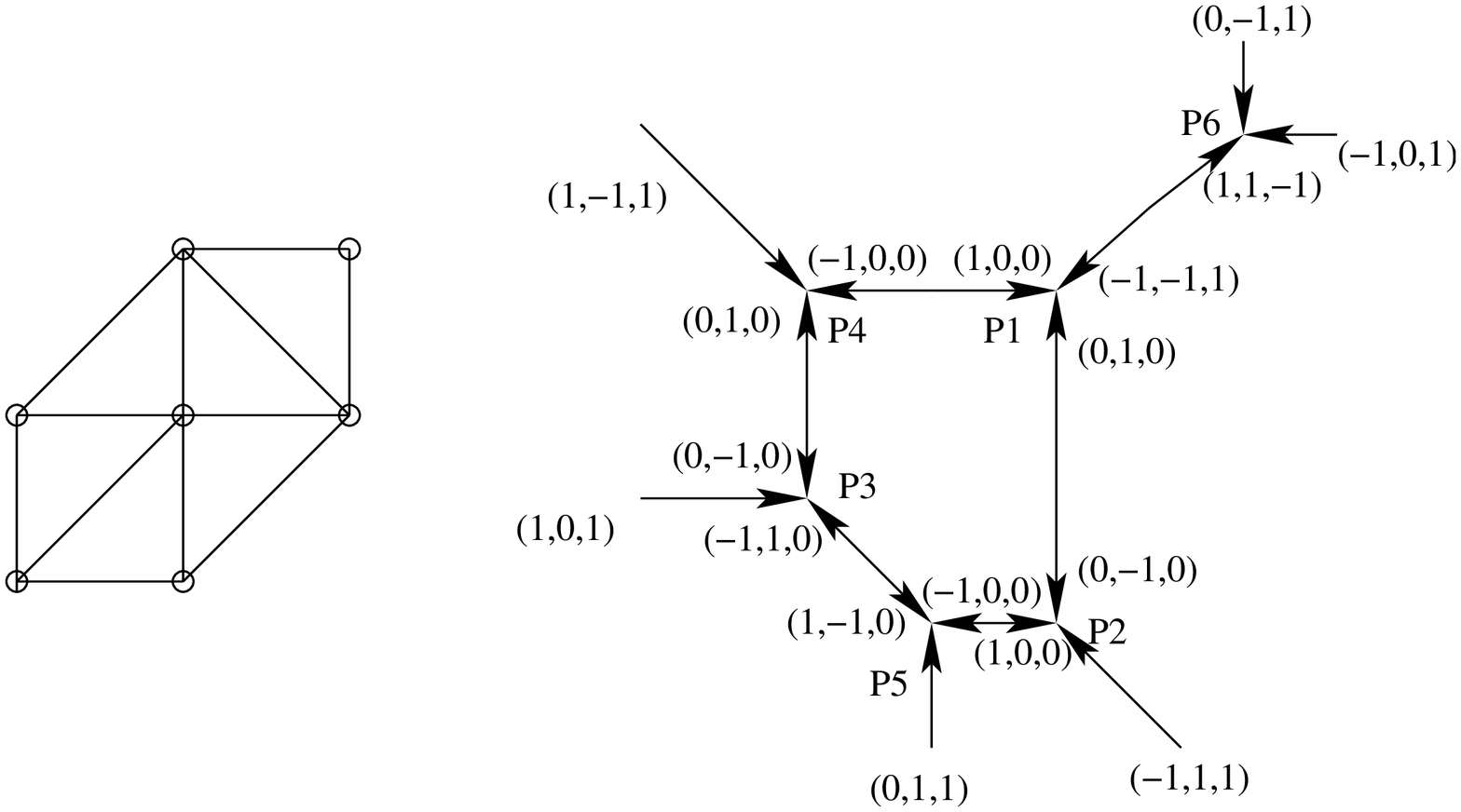}\ea \\
(c)~~dP_2 && (d)~~dP_3\\
\ea\]

For completeness, we also include the data for $F_0$ as well as 
$PdP4$, the so-called pseudo $dP_4$ surface, first introduced in
\cite{Feng:2002fv}, which is obtained from blowing up a non-generic point of
$dP_3$ so as to keep it a toric variety:\\
\[\ba{cc}
\ba{l}\epsfxsize=3in\epsfbox{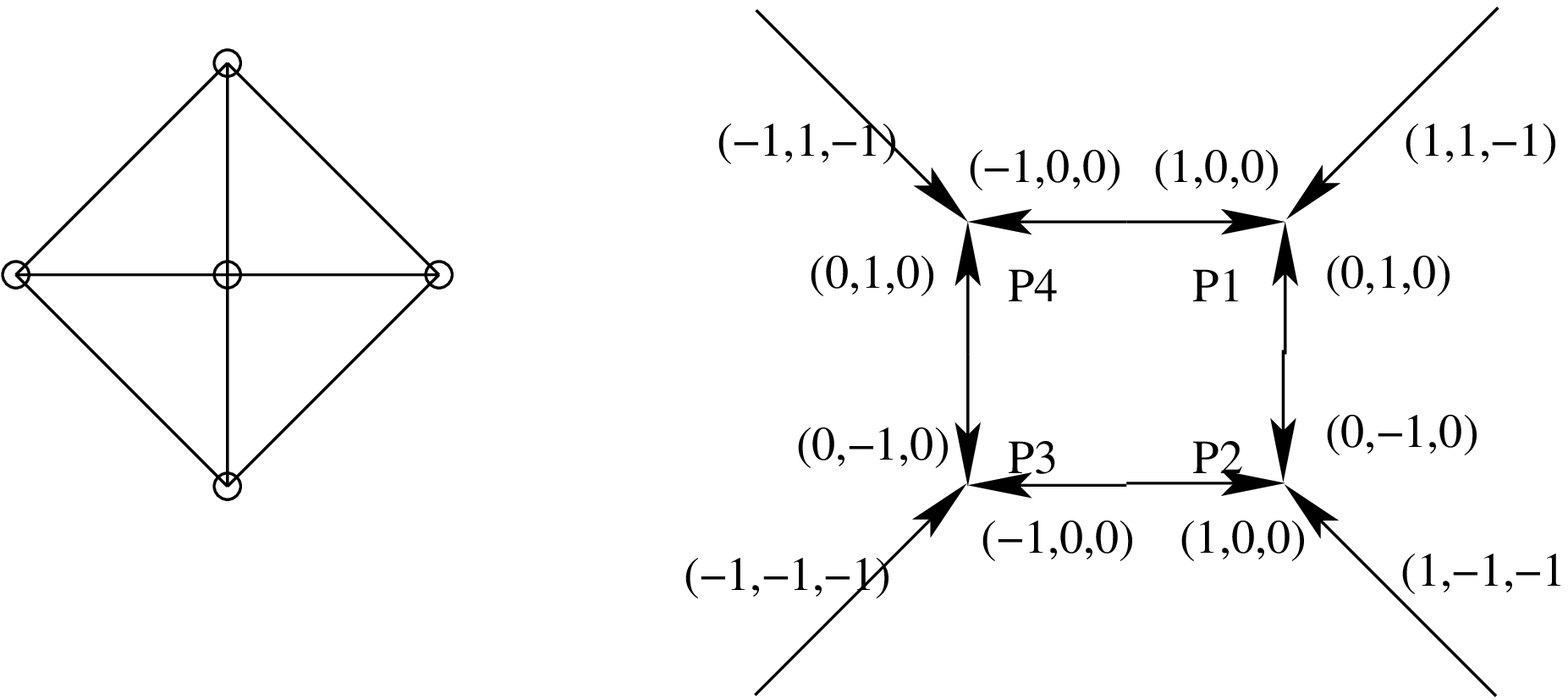}\ea&
\ba{l}\epsfxsize=3in\epsfbox{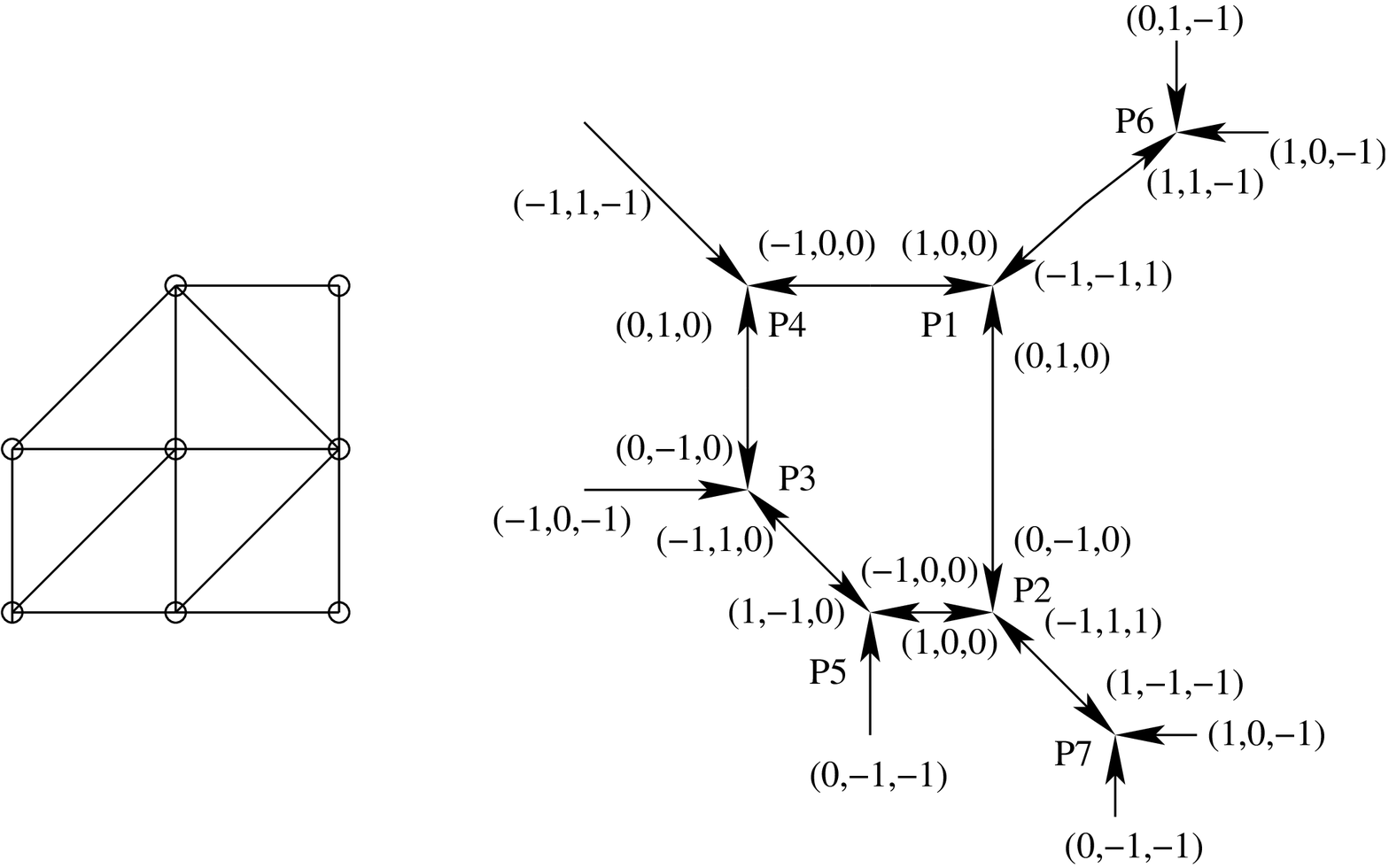}\ea\\
(a)~~F_0 & (b)~~PdP_4
\ea\]

\subsubsection{A General Formula for $dP_n$}
We can see that taking the limit $x=y \to 1$ to relax the refinement
in \eref{f-dP0123} the expressions become very simple. In other
words, we neglect the two $U(1)$ charges carried by $x,y$ and keep
only the $U(1)$ charge carried by $t=z$ (note that this is not the
R-charge but a linear combination of the three $U(1)$ charges).  
The result counts the
single-trace GIO's of a given total degree: \beq\label{ft-dp0123}
\ba{ccc} f(t; dP_0) = \frac{1+7t+t^2}{(1-t)^3}, &&
f(t; dP_1) = f(t; F_0) = \frac{1+6t+t^2}{(1-t)^3}, \\
f(t; dP_2) = \frac{1+5t+t^2}{(1-t)^3}, &&
f(t; dP_3) = \frac{1+4t+t^2}{(1-t)^3} \ .
\ea\eeq
We shall see in the
next section what it means to set $x,y$ to 1 and how all this relates
to projecting 3-dimensional toric diagrams to 2-dimensions and to
dimers. For the mean time,
observing the pattern \eref{ft-dp0123}
for the above 4 members of the del Pezzo
family, we propose the following general expression for the generating
function:
\beq\label{f-dpn}
f(t)^{(n)} := f_\infty(t; dP_n) = \frac{1+(7-n)t+t^2}{(1-t)^3},
\qquad n=0,\ldots,8 \ .
\eeq
We remark that the result for $F_0$ is the same as $dP_1$. This is
not surprising because they both, when having 1 more generic point
blown-up, become $dP_2$. Also, $dP_0$ is a Calabi-Yau over $\IP^2$, it
is in fact simply the orbifold $\IC^3 / \IZ_3$ which we will encounter
again in \sref{s:plogsyz}. Furthermore, setting $n=4$ gives agreement
with the recent $(P)dP_4$ \cite{Feng:2002fv}
result of Eq. 5.29 of \cite{Butti:2006nk}. 
Indeed, we shall revisit the del Pezzo
family, and give full credence to \eref{f-dpn} in \sref{s:dP-re}.

%
%------------
%
\section{Dimers, Toric Diagrams and Projections}\label{s:dimer}
\setall
By now we have seen the Poincar\'e series $f_\infty$
in full action in counting single-trace GIO's. Before proceeding to
finding the generating function $g_\infty$ for the multi-trace case,
let us take a brief but important interlude in how the counting in $f$
is pictorially realised for toric varieties. In due course, we shall
see how the invariants emerge in slices of the 3-dimensional toric
cones and how such projections relate to dimers and 2-dimensional
tilings. Indeed, it is these observations which initiated our
original interest in this problem of counting GIO's.

\subsection{Example: Dimers and Lattices for $\IC^3$}
\begin{figure}[h]
$\ba{ccc}
\ba{l}\epsfxsize=5cm\epsfbox{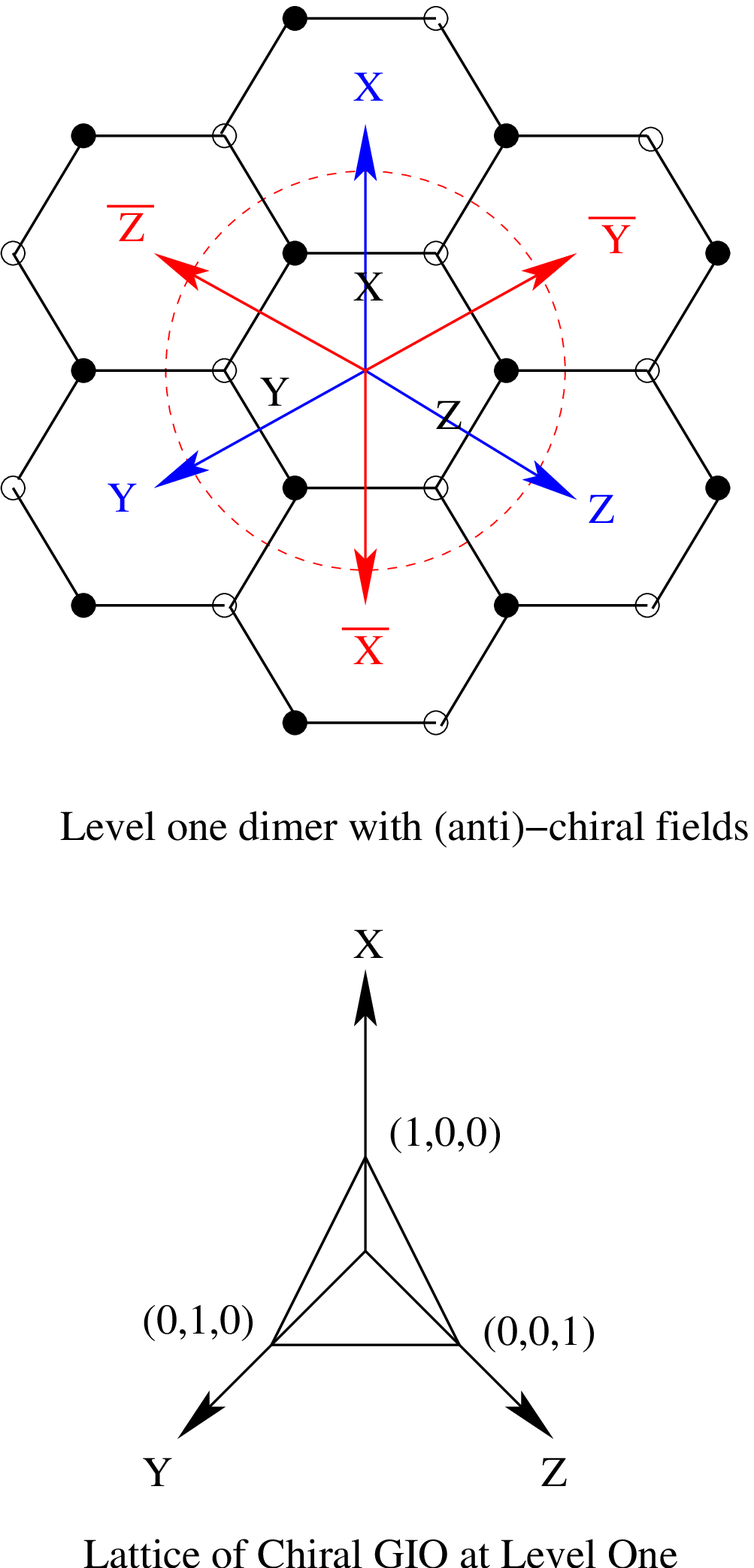}\ea &
\ba{l}\epsfxsize=4.5cm\epsfbox{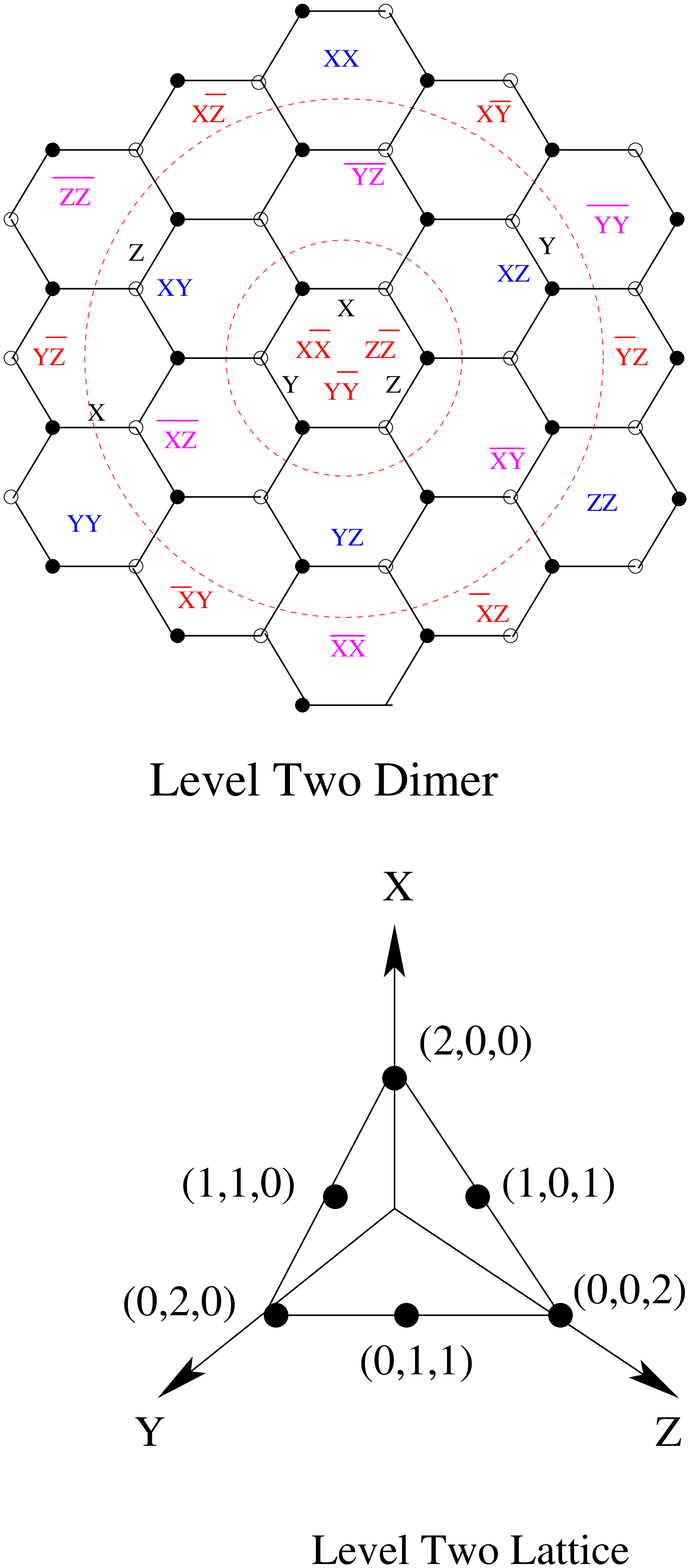}\ea &
\ba{l}\epsfxsize=5cm\epsfbox{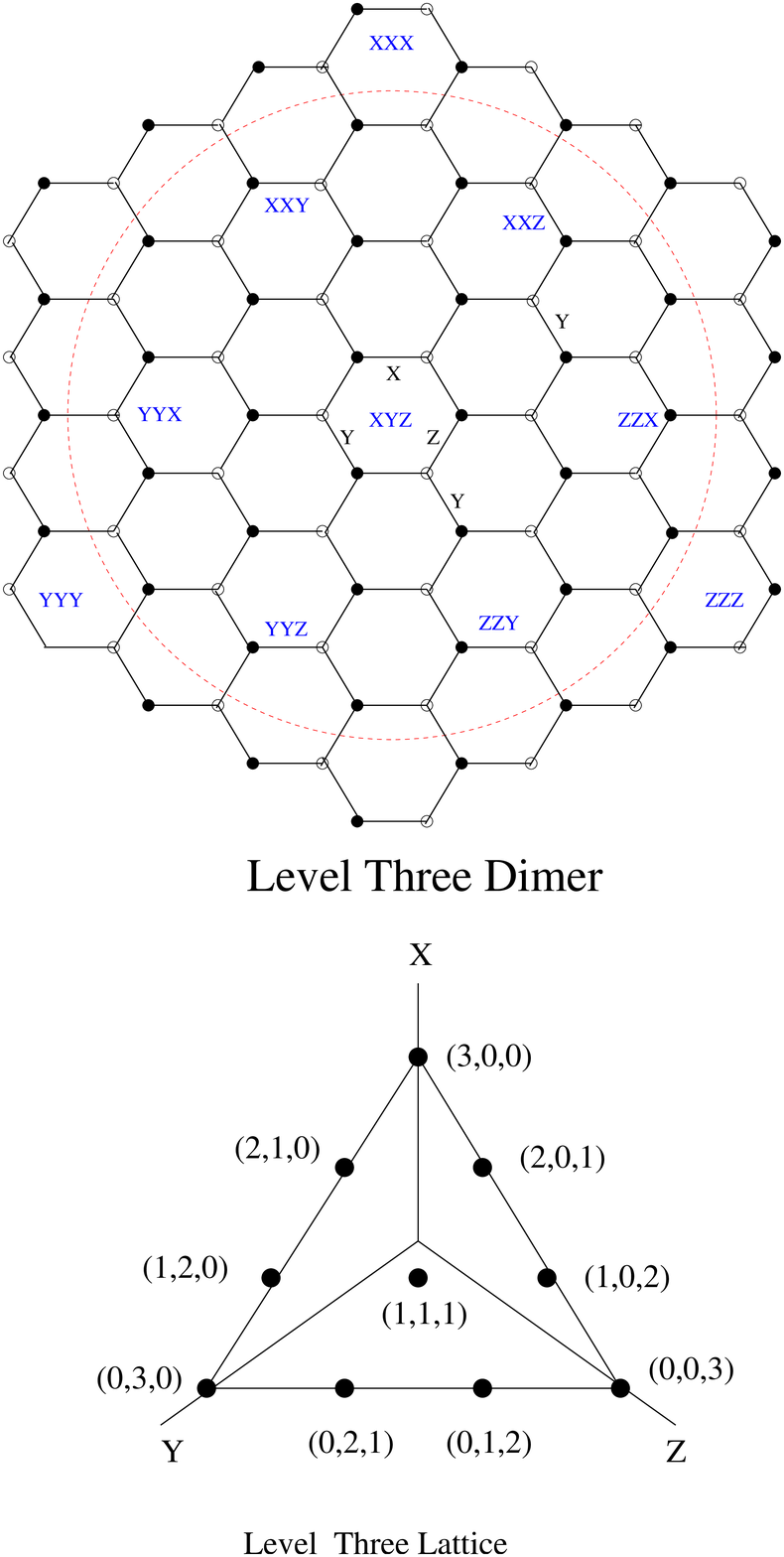}\ea
\ea$
\caption{{\sf The Dimer configurations and the lattice structure of
    GIO's for $\IC^3$, exhibited at the first 3 levels. We have drawn
    some mixed chiral-antichiral GIO's as well for illustration, 
    but the ones of
    our concern, viz., the chiral ones, are drawn in blue.}}
\label{f:C3dimer}
\end{figure}

We begin by discussing the simplest toric CY 3-fold, $\IC^3$, which
was first mentioned in \sref{s:c3eg} and then in \sref{s:toric}. 
Let us see how to represent the chiral ring in the dimer diagram of
$\IC^3$. The dimer for $\IC^3$ is well-known \cite{Hanany:2005ve} 
and is drawn in \fref{f:C3dimer}.
There is only one gauge group and it is represented by a
hexagon. We recall the fundamental fact that in a dimer, the polygonal
faces correspond to gauge groups, edges, (perpendicular) to fields and
nodes, to superpotential terms. Thus, a BPS GIO in the chiral ring
\footnote{
Since we are studying BPS mesons, we may emphasize the relationship
between the concept of the ``extremal BPS meson'' 
(introduced in \cite{Benvenuti:2005cz})
and that of the ``zig-zag path'' in
\cite{Hanany:2005ss,Feng:2005gw,Hanany:2006nm}.}
can be thought of as a
path from the origin to a polygon, and we shall show below that it is
in fact path-independent.
Since we consider only chiral operators, we represent
the (holomorphic) operator by oriented lines crossing edges,
such that
when a line crosses an edge, the black vertex is to its left
(recall that the coloring convention in a dimer has orientation built
in).
There are three holomorphic fields denoted by $X,Y,Z$.
As mentioned before, the $F$-term relations here make these three
operators mutually commutative.

Now, we discuss the chiral GIO's in detail.
We shall do so according to the number of {\bf levels}.
Here, we define level to mean the number of $X,Y,Z$ fields
inside the chiral operators. This was what we meant by {\it degree} in
the aforementioned generating function. For clarity, we have enclosed
each level with a dotted red circle in the diagram.
At level 1, there are only 3,
given by ${\tr}(X)$, ${\tr}(Y)$ and
${\rm Tr}(Z)$. This has been shown in level one of \fref{f:C3dimer}. 
In the
figure we have given also the 3 anti-chiral operators ${\rm
Tr}(\bar{X})$, ${\rm Tr}(\bar{Y})$ and ${\rm Tr}(\bar{Z})$ for
  reference. Henceforth, we shall use {\it blue} to denote the GIO's
  in which we are interested, viz., the chiral single-traces ones. The
  3 here, of course, correspond to the $3t$ term in \eref{C3orb}.

Next, let us move to level 2 of \fref{f:C3dimer}. This
time, we can cross two edges, as shown by the second red circle. A
few remarks are at hand. First, for the hexagon denoted by the blue $XY$
(which is inside the chiral ring), we have two paths to reach from the
center. One is from the center to hexagon $\overline{YZ}$ then to $XY$.
Another one is from the  center to $\overline{XZ}$ then to $XY$. The key
point is that these two paths give the same element ${\rm Tr}(XY)$
in the chiral ring. So, our first conclusion is that for the
chiral ring,
a GIO depends only on the starting and ending point of the path in the
dimer model and does not depend on the path itself.
This in fact is generic for every dimer model and not just for the
simple hexagonal model discussed here. See
\cite{Hanany:2005ss,Hanany:2006nm,Benvenuti:2005cz} 
for a proof of this point.

Furthermore, we have in fact drawn not only the chiral ring, but also the
anti-chiral ring and the mixed chiral-antichiral operators\footnote{
We remark that the mixed operators are protected only in $\cN=4$
because of enhanced SUSY, in generic $\cN=1$ theories the ``protected
rings'' are only the chiral or antichiral ones. 
The other 1/2-BPS protected operators like the currents do not form a
ring.
}
at level 2. For the
mixed operators, it is easy to see that now the path in the dimer does
matter. We start from the center, go up and then go down and get
$\bar{X}X$. Similarly we can go southwest to get $\bar{Y}Y$ and
southeast to get $\bar{Z}Z$. This means that we should put all three
$\bar{X}X$, $\bar{Y}Y$ and $\bar{Z}Z$ in the center hexagon. In
another word, the one-to-one correspondence we found for chiral or
anti-chiral ring is lost. Because of this
complexity we will not discuss mixed operators further in this
paper (the anti-chiral ring is isomorphic to the chiral ring and need
not be addressed separately).

Focusing on only the chiral ring we can see that there are 6 (chiral)
GIO's at level 2, corresponding to the term $6t^2$ in \eref{C3orb}.
Also, The result of level 3 is given at the right of
\fref{f:C3dimer}. Here,
we give only the chiral ring operators at proper hexagons in this
figure. There is a total of 10 as shown, corresponding to the
$10t^3$ in \eref{C3orb}.
\subsubsection{Lattice Structure and Planar Slices}
We thus conclude that:
\qq{
A (chiral) GIO's at level $n$ corresponds to a 
polygon in the dimer, which is a chiral-distance $n$ away from the
center.
}
In the above, a chiral-distance is measured by segments of only chiral
operators, i.e., black vertices to the left. We have drawn these
chiral GIO's in blue in \fref{f:C3dimer}.
With this one-to-one
correspondence between chiral GIO's at a given level and hexagons in
the dimer diagram, we can see that in fact we have a lattice
structure in $\IR^3$. Each integer lattice point $(a,b,c)$ with
$a,b,c\geq 0$ corresponds to a chiral operator. The level of this
operator is given by $(a+b+c)$. In other words, level $n$ is given
by the plane perpendicular to vector $(1,1,1)$ and has distance $n$
from the origin. It is interesting to notice that the vector $(1,1,1)$
is the Reeb vector of $\IC^3$.
Thus the degree we are counting is
exactly the R-charge. We have drawn these plane slices for each
level in our figure as well in \fref{f:C3dimer}.

This lattice is something with which we are familiar! It is nothing
other than the dual toric cone for $\IC^3$. Indeed, the definition
of a toric variety is that it is the affine spectrum of the ring of
monomials obtained from raising the coordinates to the powers of the
lattice generators, i.e., the invariant monomials. This is what we
are doing above. Level 1 gives the monomials which are obtained from
the lattice generators of the cone; level 2 gives the monomials
obtained from the toric cone intersected with the (non-primitive)
lattice points one further step away, etc.

\subsection{Example: Dimers and Lattices for the Conifold}

\begin{figure}[h]
$\ba{ccc}
(a) && \ba{l}\epsfxsize=4in\epsfbox{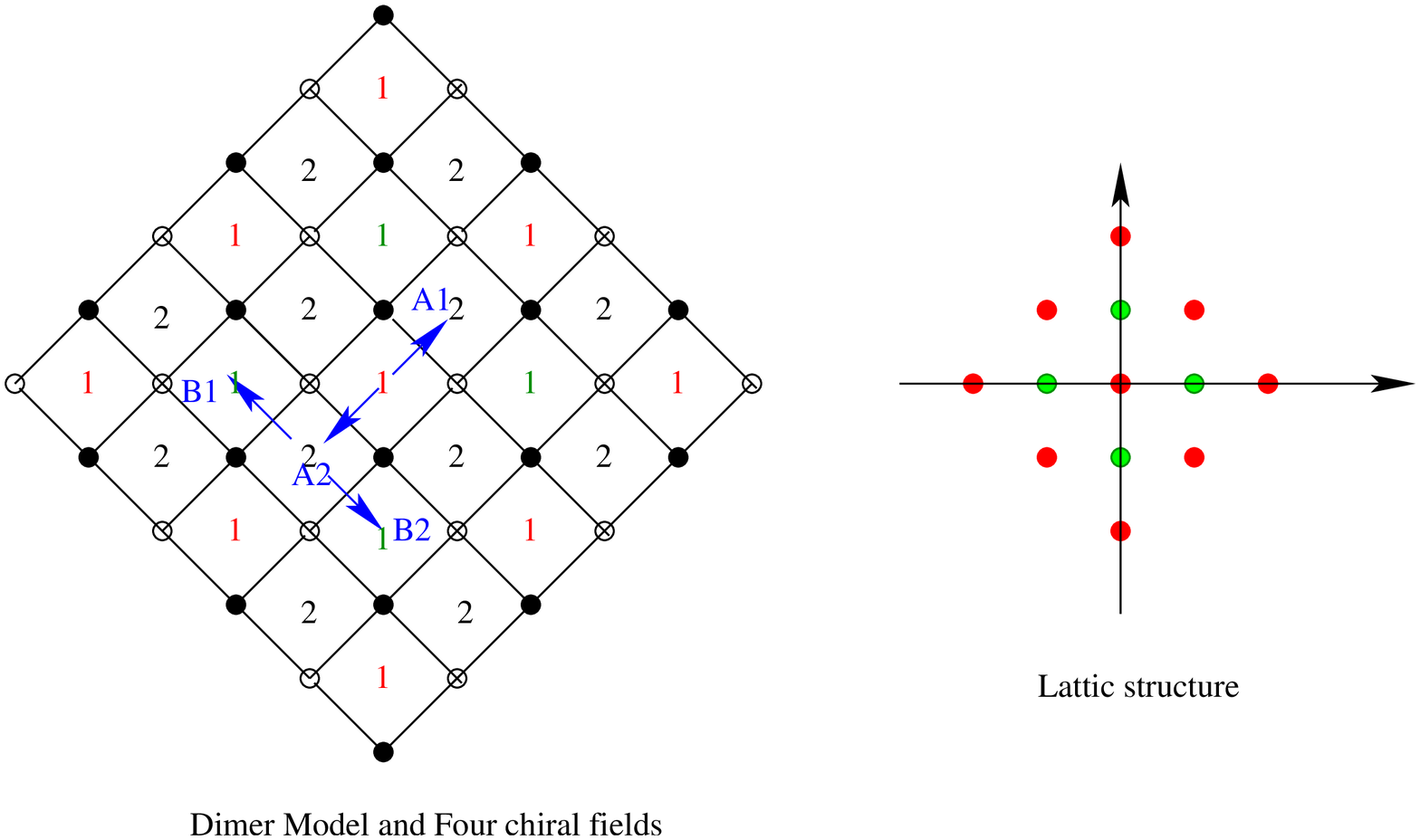}\ea \\
(b) && \ba{l}\epsfxsize=6in\epsfbox{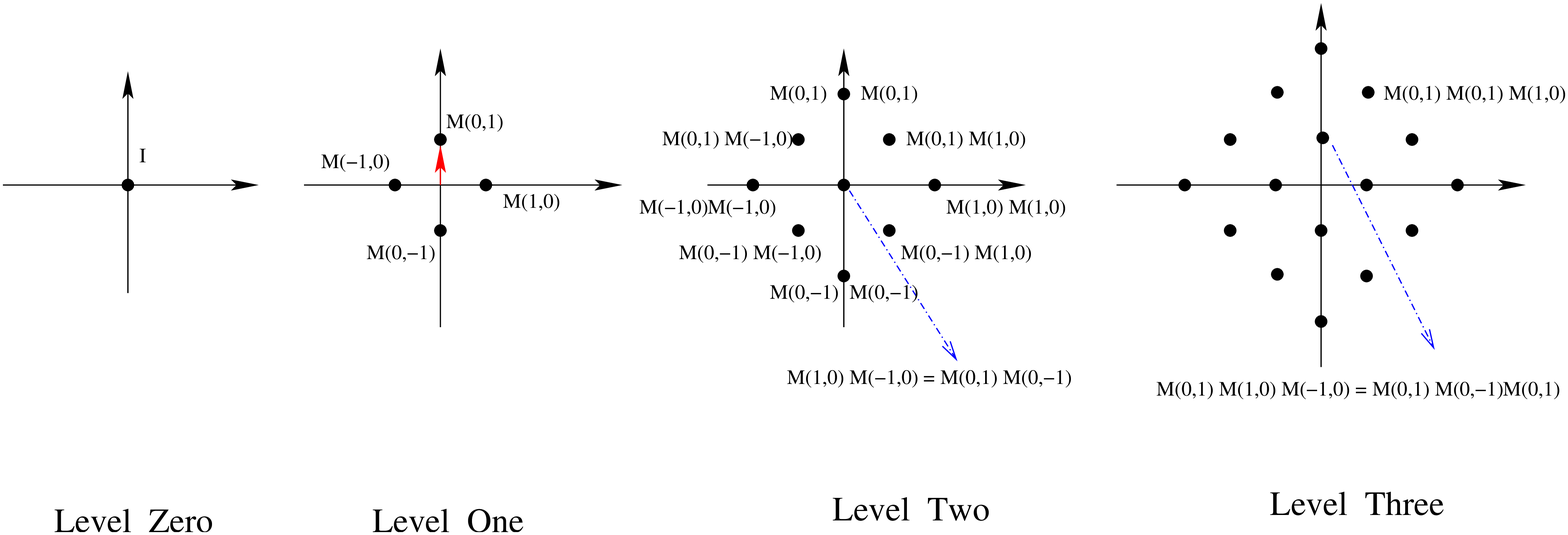}\ea
\ea$
\caption{{\sf (a) The dimer configuration for the conifold and the
    dual lattice structure of GIO's; (b) In more detail, the actual
    operators (with built-in relations) corresponding to the lattice
    points at the first 3 levels. }}
\label{f:conidimer}
\end{figure}

Next, let us discuss the conifold. Here, we will see explicitly 
how we must count the GIO's up to relations from F-terms, as was
mentioned in the introduction.
The toric diagram was given in
\fref{f:coniflop} in \sref{s:toric}. The dimer model is the brane
diamond \cite{Aganagic:1999fe} drawn on $T^2$ \cite{Hanany:2005ve} 
and is given in \fref{f:conidimer}.
There are two gauge groups so there will be two
types of polygons which are labeled 1 and 2. It is easy to see that
we can locate these 2 gauge groups at lattice points. More
explicitly, if we draw the lifting of $T^2$ in $\IR^2$, we can
identify an integer lattice point $(a,b)$ to gauge group $1$ if
$a,b$ are integers or gauge group $2$ if $a,b$ are half-integers.
Now, since we are considering the single-trace mesonic GIO's, we can
neglect gauge group $2$ and consider the holomorphic paths
connecting different lattice points of gauge group $1$, i.e.,
integer lattice points in 2-dimensions.

In this gauge theory, there are four bi-fundamental fields $A_1,
A_2$ and $B_1, B_2$. We define the following operators which are in
the adjoint representation of gauge group $1$: 
\beq 
M_{0,1}= A_1
B_1,\qquad M_{1,0}=A_1 B_2,\qquad M_{-1,0}=A_2 B_1,\qquad
M_{0,-1}=A_2 B_2 \ . 
\eeq 
It is easy to check that the $F$-term
relations tell us that all four $M_{ij}$ commute and obey one
non-trivial relation: 
\beq\label{coni-rel} 
M_{0,1} M_{0,-1}= M_{1,0}M_{-1,0} \ . 
\eeq

We can map the above quantities into the dimer model.
As we have shown above, the
dimer model can be mapped to a 2-dimensional integer lattice. The
operator $M_{0,1}$ can be mapped to vector $(0,1)$ so we can use it
to connect points $(0,0)$ and $(0,1)$. Similarly, $M_{1,0}, M_{-1,0},
M_{0,-1}$ map to vectors $(1,0),(-1,0), (0,-1)$, respectively.
Using this mapping, a
single-trace GIO is mapped to a path connecting point $(0,0)$ to
$(a,b)$ using the above four vectors. The non-trivial relation
\eref{coni-rel} is nothing, but the statement that after
following consecutively vectors $(0,1)$ and $(0,-1)$
(or $(1,0)$ and $(-1,0)$) we come back to the starting point.

Using this picture we can see the lattice structure of holomorphic
GIO's. For level $0$, it is the origin $(0,0)$ and corresponds to the
identity operator. For level one, we can use only one $M_{i,j}$ to
connect $(0,0)$ to nearby lattice points. Thus we have four of them
${\rm Tr}(M_{1,0})$, ${\rm Tr}(M_{-1,0})$, ${\rm Tr}(M_{0,1})$, and
${\rm Tr}(M_{0,-1})$.
For level two, we need to use two $M_{i,j}$ operators. It is easy to
get to
lattice points $(\pm 2,0), (\pm 1,\pm 1)$ as well as $(0,0)$. For
$(0,0)$ we have two ways ${\rm Tr}(M(0,1)M(0,-1))$ or ${\rm
Tr}(M(1,0)M(-1,0))$. But by relation \eref{coni-rel} they are the same
so we should count only once. Similarly we can draw the level three
lattice diagram as shown in \fref{f:conidimer}.

\subsubsection{Planar Slices and Lattices}
Now let us find the 3-dimensional box which projects to the above
2-dimensional picture. The vectors $(0,0,1)$, $(1,0,1)$,
$(0,1,1)$ and $(1,1,1)$ of the toric diagram of $\cC$ generates an
integral cone; we can find the
generators of the dual cone to be $v_1=(1,0,0)$, $v_2=(0,1,0)$,
$v_3=(0,-1,1)$ and $v_4=(-1,0,1)$. By definition, 
a lattice point in the dual cone is given by positive integer 
linear combinations of these four
vectors. It is special in our case that these four generators $v_i$
have their endpoints co-planar\footnote{Note that though the toric
  diagram always has its vectors in a plane, as guaranteed by the CY
  condition, the dual cone is not so guaranteed.}.
It is easy to find
 the vector $u$ orthogonal to the plane generated by
$v_i$ as $u=(1,1,2)$. In fact, in this case, the vector $u$ is
 precisely the Reeb
vector, so the level we are counting is also the R-charge\footnote{We
  remark that our Reeb vector differs in convention from that of
  \cite{Martelli:2006yb}. Our $(a,b,c)$ is their $(c,a,b)$.}.

Now, we can see how this 3-dimensional lattice generated by $v_i$
projects to the 2-dimensional lattice. For level one, it is given by
four $v_i$, since all of them have $v_i\cdot (1,1,2)=1$. For the
level two, we need to find vectors $(x,y,z)$ such that (1) $(x,y,z)=
\sum_{i=1}^4 a_i v_i$ with $a_i\geq 0$ and integer; (2)
$(x,y,z)\cdot (1,1,2)=2$ which gives $x+y+2z=2$.  From these
conditions, we find the following $9$ points: 
\beq \{
(2,0,0),(0,2,0),(0,-2,2),(-2,0,2),(1,1,0),
(1,-1,1),(0,0,1),(-1,-1,2), (-1,-1,2)\} 
\eeq 
which is exactly what
we find in the dimer model. In general level $n$ should have $(n+1)^2$
points.

Now let us check this using our Poincar\'e series, which from
\eref{coni1} is 
\beq\label{Pxyz-coni}
P(x,y,z;~\cC) =
- \frac{x\,y\,\left( -1 + z \right) }
    {\left( -1 + x \right) \,\left( -1 + y \right) \,
      \left( x - z \right) \,\left( y - z \right) } \ .
\eeq 
To count the level, notice that the Reeb
vector is $u=(1,1,2)$, which means that the R-charges of
$x,y,z$ are $1,1,2$ respectively. In other words, we should replace
$x\to q, y\to q, z\to q^2$, yielding
\beq\label{P-coni}
P(q;~\cC) =
{(1+q)\over(1-q)^3} = \sum_{n=0}^{\infty} (n+1)^2 q^n \ ;
\eeq 
whereby giving us the required $(n+1)^2$ counting!

In fact we can do better than that. Let us do the following
replacement $x\to x q$, $y\to y q$ and $z\to q^2$. The expression is
changed to
\beq\label{f-xyq-coni}
{xy (1-q^2)\over (1-q x)(1-q y) (q-x)(q-y)} = 1+
q(x+y+{1\over x}+{1\over y})+ \ldots \ .
\eeq
Comparing this with toric data we can see that $x,y$ represent the
Cartan weight of $SU(2)_L\times SU(2)_R$ global symmetry for the
conifold. More explicitly, for the two $U(1)\times U(1)$, $x,y$ carry
 the charge of $U(1)_x=U(1)_L+U(1)_R$ and $U(1)_y= U(1)_L-U(1)_R$.
To see this let us consider, for  example, $M_{0,1}=A_1B_1$. Because the 
$(U(1)_L, U(1)_R)$ charge of $A_1$ and $B_1$ is $(\frac12,0)$ and
$(0,\frac12)$, we get immediately the $(U(1)_x, U(1)_y)$ charge
$(1,0)$, i.e., the term $x$.
Similarly, the terms
 $y,{1 \over x},{1 \over y}$ correspond to the operators
$M_{0,1}, M_{-1,0}, M_{0,-1}$, respectively.

%
%-------------- MULTI-TRACE
%
\section{Counting Gauge Invariants: Plethystics, Multi-Trace and
  Syzygies}\label{s:multi}
\setall
We have now seen the generating function $f$ which counts single-trace
GIO's of a given choice of global charges
for 3 large families of CY threefold
singularities. What about the multi-trace GIO's? These are products of
combinations of single-traces. We have called the generating function
for counting these, $g$. We shall now see how $g$ can be obtained
from $f$ using some nice combinatorics. We shall then see how the
function which relates $f$ and $g$ has some remarkable geometrical
properties as well.

\subsection{The Plethystic Exponential: From Single to Multi-Trace}
Recall that in the above $f$ should really be $f_\infty$ bacause we
have taken the large $N$ limit. Similarly, the quantity $g$ we desire
is really $g_\infty$.
\comment{
An immediate observation is that
\beq\label{g1finf}
g_1 = f_\infty \ .
\eeq
Indeed, this is true because at $N=1$ the matrix GIO is just a complex
number and there is no distinction between single- and multi-trace.
Next, we recall the matrix identity
\beq
\det(\II-M) = \exp(\tr(\log(\II-M))) =
\exp\left( \sum_{k=1}^\infty \frac{\tr(M^k)}{k} \right) \ .
\eeq
}
Now, we showed in \sref{s:c3eg} that for $\IC^3$, the relation between $f$
and $g$ is that of the {\bf plethystic exponential}, $PE$
(q.v.~\cite{Labastida:2001ts,plethy}). This in
fact holds in general:
\beq\label{pexp}
g(t) = PE[f(t)] := \exp\left(
\sum_{k=1}^\infty \frac{f(t^k) - f(0)}{k} \right) \ .
\eeq
Indeed, recalling \eref{g1finf}, we summarise the following relations,
with the subscripts restored:
\beq\label{summary-fg}
g_1 = f_\infty, \qquad
f_\infty(t) = PE[f_1(t)], \qquad
g_\infty(t) = PE[g_1(t)] = PE[PE[f_1(t)]] \ .
\eeq
We remark that, even for a list of variables $t_{i=1,\ldots,n}$, which
are used in the refinement of counting discussed above, the
expressions in
\eref{pexp} and \eref{summary-fg} still hold, with obvious
replacement. Namely,
\beq
g(t_1,\ldots,t_n) = PE[f(t_1,\ldots,t_n)] 
:= \exp\left(\sum\limits_{k=1}^\infty \frac{f(t_1^k,\ldots,t_n^k) -
  f(0,\ldots,0)}{k} \right) \ .
\eeq
We can derive the statement \eref{pexp} explicitly by
series-expansion. Let
\beq
f(t) = \sum\limits_{n=0}^\infty a_n t^n
\eeq
be the Taylor expansion of the Poincar\'e series $f_\infty=f(t)$.
Thus, $a_n$ is the
number of independent invariants at (total) degree $n$.
Then, \eref{pexp} gives us
\bean\nn
PE[f(t)] &=& \exp\left(
\sum\limits_{n=0}^\infty a_n \sum\limits_{k=1}^\infty \frac{t^{nk}}{k}
- a_0 \sum_{k=1}^\infty\frac1k \right)
\\ \nn
&=& \exp\left(
- \sum\limits_{n=0}^\infty a_n \log(1-t^n)
- a_0 \sum_{k=1}^\infty\frac1k
\right) \ . \\
\eean
We see therefore that the $f(0)$ term precisely regularises the sum
and we obtain
\beq\label{eulerPE}
PE[f(t)] = \exp\left( - \sum\limits_{n=1}^\infty a_n \log(1-t^n)
\right) = \frac{1}{\prod\limits_{n=1}^\infty (1-t^n)^{a_n}} \ .
\eeq
This expression is now in the standard Euler product form.
Upon expansion of $PE[f(t)]$,
we would see that the coefficient for $t^m$ is the
number of ways of partitioning $m$, each weighted by $a_n$. This is
precisely our required counting, i.e., the
number of multi-trace GIO's at degree $m$. Hence, $g(t) = PE[f(t)]$.

We have thus solved problems (1) and (2) posed in the introduction and
have the generating functions $f$ and $g$ for large $N$. In fact, as
before, we can refine our counting. In the above, we had a single
variable $t$, a dummy variable associated with the total degree. Where
permitted, as discussed in \sref{s:refine}, we can have a set of
variables $t_i$, one for each $U(1)$-charge, and an associated
multi-degree for these tuples of charges. In addition, we can
introduce one more variable $\nu$, to be inserted into the summand.
One could easily see that upon expansion, the power of
$\nu$ will actually count how many single-trace operators are present
in each of the terms. In other words, for $f_\infty(t_1,\ldots,t_m) =
\sum\limits_{p_1, \ldots, p_m = 0}^\infty a_{p_1,\ldots,p_m} t_1^{p_1}
\ldots t_m^{p_m}$, we have
\beq\label{Pnu}
\ba{rcl}
\tilde{g}_\infty(t_i, \nu) &=& PE[f_\infty] = \exp\left(
\sum\limits_{k=1}^\infty
\frac{f_\infty(t_1^k,\ldots,t_m^k) \nu^k}{k}
\right) \\
&=& \left({\prod\limits_{p_1, \ldots, p_m}
(1- \nu t_1^{p_1}\ldots t_m^{p_m})^{a_{p_1,\ldots,p_m}}} \right)^{-1}
\ ;
\ea\eeq
%where the product is taken over all non-negative $p_i$ with the point
%$(p_1, \ldots, p_m)=(0,\ldots,0)$ excluded. 
note that due to the insertion of $\nu$, there is no longer a need to
regulate the sum by the subtraction of $f(0,\ldots,0)$.

\subsubsection{The Plethystic Logarithm}
The inverse function of $PE$ is also a fascinating one.
It is called the plethystic logarithm \cite{Labastida:2001ts}; 
one can in fact write it analytically:
\beq\label{plog}
f(t) = PE^{-1}(g(t))
= \sum_{k=1}^\infty  \frac{\mu(k)}{k} \log (g(t^k)) \ ,
\eeq
where $\mu(k)$ is the M\"obius function
\beq
\mu(k) = \left\{\ba{lcl}
0 & & k \mbox{ has one or more repeated prime factors}\\
1 & & k = 1\\
(-1)^n & & k \mbox{ is a product of $n$ distinct primes}
\ea\right. \ .
\eeq

As $g_\infty = PE[g_1]$, so too does one have the relation $f_\infty =
PE[f_1]$. Since our basic generating function is the Poincar\'e series
$f = f_\infty$, for which we
have had explicit results in \sref{s:single}, it is more convenient to
write
\beq
f_1 = PE^{-1}(f_\infty) \ .
\eeq
One may ask what this function $f_1$, which we briefly encountered in 
\eref{f1C3}, signifies.
It has a remarkable geometrical property!
\qq{
The plethystic logarithm of the
Poincar\'e series, is a generating series
for the relations and syzygies of the variety!
}
We exemplify this statement
with our familiar example of the Valentiner group
from \sref{s:delta27} in the next subsection.
\subsubsection{Plethystic Logarithm and Syzygies}\label{s:plogsyz}
Using \eref{plog}, and recalling the Poincar\'e (Molien) series $f$
for $\Delta(27)$ from \eref{Mdel27}, we see that
\beq\label{del27plog}
f_1 = PE^{-1}\left( \frac{-1 + t^3 - t^6}{{\left( -1 + t^3 \right)
  }^3}\right) = 2 t^3 + t^6 + t^9 - t^{18} \ .
\eeq
The RHS terminates and is a polynomial!
It is to be interpreted thus: there are 2 degree 3 invariants, 1
degree 6 and 1 degree 9 invariant, these 4 invariants obey a single
relation of total degree 18. Upon inspecting \eref{invdelta27} and
\eref{eqdelta27}, we see that this is indeed the definition of
$\IC^3/\Delta(27)$ as a variety!

Now, $\IC^3/\Delta(27)$, as a hypersurface in $\IC^4$, is a complete
intersection affine variety (i.e., the number of equations is equal to
the codimension of the variety in the embedding space). How does the
above work for non-complete intersections? We have an example readily
available: the famous $\IC^3/\IZ_3 =
\cO_{\IP^2}(-3)$ orbifold. In fact, being an Abelian orbifold, this is
also toric and furthermore, it is also $dP_0$, being a cone over
$\IP^2$. So we have 3
ways to compute its Molien series from \sref{s:single}. Let us use the
Molien series. The action is $(x,y,z) \to \omega_3(x,y,z)$
and we immediately get
\beq
f_\infty(t) =
M(t; \IZ_3) =  \frac{1 + 7\,t^3 + t^6}{{\left( 1 - t^3 \right) }^3} \
,
\eeq
whereby
\beq\label{plogZ3}
f_1(t) = PE^{-1}[f_\infty(t)] =
10\,t^3 - 27\,t^6 + 105\,t^9 - 540\,t^{12} +
  3024\,t^{15} - 17325\,t^{18} + \cO(t^{21}) \ .
\eeq
This is again in accordance with known facts! The equation for this
orbifold is 27 quadrics in $\IC^{10}$, i.e., 10 degree 3 invariants
satisfying 27 relations of degree $2 \times 3 = 6$ 
(q.v.~\cite{Yau,grob}).
We can determine these as follows. The 10 invariants are
\beq\label{c3z3inv}
y_{1,\ldots,10} =
\left\{{x}^3 , \  {x}^2y , \  x{y}^2, \  {y}^3, \  {x}^2z , \
  xyz, \  {y}^2 z, \   x{z}^2, \  y{z}^2, \ {z}^3
\right\} \ ,
\eeq
obeying the 27 quadrics
\beq\label{c3z3syz}
\ba{l}
\{ {y_2}^2 - y_1y_3, \,y_2y_3 - y_1y_4, \,
  {y_3}^2 - y_2y_4, \,y_2y_{5} - y_1y_{6}, \,
  y_3y_{5} - y_1y_{7}, \,y_2y_{6} - y_1y_{7}, \, \\
  y_4y_{5} - y_2y_{7}, \,y_3y_{6} - y_2y_{7}, \,
  y_4y_{6} - y_3y_{7}, \,{y_{5}}^2 - y_1y_{8}, \,
  y_{5}y_{6} - y_1y_{9}, \,y_2y_{8} - y_1y_{9}, \,\\
  {y_{6}}^2 - y_2y_{9}, \,y_{5}y_{7} - y_2y_{9}, \,
  y_3y_{8} - y_2y_{9}, \,y_{6}y_{7} - y_3y_{9}, \,
  y_4y_{8} - y_3y_{9}, \,{y_{7}}^2 - y_4y_{9}, \,\\
  y_{5}y_{8} - y_1y_{10}, \,y_{6}y_{8} - y_2y_{10}, \,
  y_{5}y_{9} - y_2y_{10}, \,y_{7}y_{8} - y_3y_{10}, \,
  y_{6}y_{9} - y_3y_{10}, \,y_{7}y_{9} - y_4y_{10}, \, \\
  {y_{8}}^2 - y_{5}y_{10}, \,y_{8}y_{9} - y_{6}y_{10}, \,
  {y_{9}}^2 - y_{7}y_{10} \} \ .
\ea
\eeq
Therefore, \eref{c3z3inv} are the 10 primitive invariants of degree 3,
obeying 27 syzygies of degree 6 given by \eref{c3z3syz}. According to
our rule, this should read $10t^3 - 27 t^6$. These are precisely the
first two terms of \eref{plogZ3}! Indeed, because we no longer have a
complete intersection, the plethystic logarithm of the Poincar\'e
series is not polynomial and continues ad infinitum.
What about the 105 and higher terms then, do they mean anything?
We will explain this in \sref{s:syn}.

As a final example of the more subtle case of
non-complete-intersection varieties, let us
take the $\IC^3 / \IZ_5$ orbifold, with action $(x,y,z)
\to (\omega_5 x, \omega_5^2 y, \omega_5^2 z)$. We obtain:
\bea\nn
M(t; \IZ_5) &=&
\frac{-1 + t - 3\,t^3 + t^4 - 3\,t^5 + t^7 - t^8}
  {{\left( -1 + t \right) }^3\,
    {\left( 1 + t + t^2 + t^3 + t^4 \right) }^2} =
1 + 3\,t^3 + 2\,t^4 + 7\,t^5 + 5\,t^6 + 4\,t^7 +
  11\,t^8 \\
&& + 9\,t^9 + 18\,t^{10} + 15\,t^{11} +
  13\,t^{12} + 24\,t^{13} + 21\,t^{14} + 34\,t^{15} +
  {\cO(t)}^{16} \ ,
\eea
giving us
\beq\label{plogZ5}
PE^{-1}[f_\infty(t)] =
3\,t^3 + 2\,t^4 + 7\,t^5 - t^6 - 2\,t^7 - 13\,t^8 -
  12\,t^9 - 14\,t^{10} + 14\,t^{11} + 34\,t^{12} +
  72\,t^{13} + 47\,t^{14} +
  {\cO(t)}^{15} \ .
\eeq
We can find that the 3 invariants of degree 3, 2 of degree 4, 7 of
degree 5 are
\beq
y_{1,\ldots,12} :=
\{ x\,y^2,x\,y\,z,x\,z^2,x^3\,y,x^3\,z,x^5,y^5,y^4\,z,
  y^3\,z^2,y^2\,z^3,y\,z^4,z^5\} \ .
\eeq
We can easily find, using Gr\"obner algorithms,
all relations amongst these 12 invariants, giving
us 1 in degree 6 (a quadric in the 3 degree 3 invariants), 2 in degree
7, 13 in degree 8, 12 in deree 9 and 16 in degree 10. All
this is in almost in exact agreement with \eref{plogZ5}, with the only
exception being that there are 16 degree 10 relations and not 14.
Together with the issue of the higher terms in the $\IC^3/\IZ_3$ case,
we now address this discrepancy in the next subsection.

\subsection{Plethystics: A Synthetic Approach}\label{s:syn}
We have now witnessed the astounding power of plethystics in the
counting problem and have
moreover noted a tantalising fact about the geometry of the variety and
the (plethystic logarithm of) the generating function for the GIO's in
the gauge theory.
Let us now attempt to argue why some of the above examples should work.
First we note that the Poincar\'e series $f$, when finally
collected and simplified, is always a rational function. In particular
it has a denominator of
the form of products of $(1 - t^k)$ with possible repeats for $k$; the
numerator is some complicated polynomial. We
will call this the {\bf Euler form}.
The point is that the coefficient in front of the $t^k$ is always
unity and we conjecture that this is a property of the Poincar\'e
series of concern
\footnote{In fact, all Poincar\'e series we have encountered, 
  orbifold, toric, etc., 
  have this property. We do not have a rigorous proof of this right
  now and leave it to the mathematically inclined.}. 
\comment{I
don't have a proof of this right now but I suspect it's due to summing
over inverse elements of the group. Anyways, I haven't found a single
counter-example; I'll leave this for now. This fact is possibly useful
to the argument below.}

When we are taking the plethystic logarithm of $f$, due to the
explicit expression \eref{eulerPE}, we
are trying to solve the following algebraic problem: find integers
$b_n$ such that
\beq\label{plethmol}
f(t) = \frac{1}{\prod\limits_{n=1}^\infty (1-t^n)^{b_n}} \ ,
\eeq
where $f(t)$ is a given rational function in Euler form.
Note that $PE^{-1}[f(t)] = \sum\limits_{n=1}^\infty b_n t^n$, unlike
the Poincar\'e series herself, need not have all positive $b_n$. Because
$f(t)$ has Euler form, the denominator of
\eref{plethmol} is immediately taken care of. In other words, because
$f(t)$ has denominator in the form of products
of $(1-t^k)$, all positive values of $n$ and $b_n$ are just read
off. These are low values of $n$ and correspond, in the Molien case,
to some of the small invariants, including the primitive ones.
However, there is still a numerator in $f(t)$, often of complicated
form. This will give
negative $b_n$ contributions to the RHS of \eref{plethmol},
which correspond to the relations.

Take $\Delta(27)$ as an example. We need to find $b_n$ such that
\beq
\frac{1 - t^3 + t^6}{{\left( 1 - t^3 \right) }^3} =
\frac{\left( 1 - t^{18} \right) }{\left( 1 - t^6 \right)
      \,\left( 1 - t^9 \right) {\left( 1 - t^3 \right) }^2}
= \frac{1}{\prod\limits_{n=1}^\infty (1-t^n)^{b_n}} \ ,
\eeq
where we have used the identity
\beq\label{id-del27}
\frac{\left( 1 - t^3 \right) \,
    \left( 1 - t^{18} \right) }{\left( 1 - t^6 \right)
      \,\left( 1 - t^9 \right)} = 1 - t^3 + t^6 \ .
\eeq
This rational
identity is crucial and expresses even the numerator of $f$
into Euler form. Now we can read out the solution: the denominator
contributes terms $+2t^3$, $+t^6$ and $+t^9$ while the numerator
contributes $-t^{18}$. Thus $PE^{-1}[M(t)] = 2 t^3 + t^6 + t^9 -
t^{18}$. In other words, there should be 2 degree 3 invariants, 1 each
of degrees 6 and 9, obeying a single relation of degree 18. The fact
that the numerator can be factorised into (finite polynomial) Euler
form dictates that the plethystic logarithm of $f$ terminates in
series expansion as was seen in \eref{del27plog}.
The moral
of the story is that
\qq{
Findings relations in this language corresponds to finding appropriate
factorisations of the numerator into Euler form.
}

Of course, not all Poincar\'e series have polynomial plethystic
logarithms. This is just the statement that not all polynomials afford
identities of the type \eref{id-del27}.
In general, the product on the RHS of \eref{plethmol}
must be infinite to
accommodate those which cannot be put into finite Euler form. These
correspond to non-complete intersection varieties.
Take $\IC^3/\IZ_3$, we have
\beq
M(t; \IC_3) =
\frac{1 + 7\,t^3 + t^6}{{\left(1 - t^3 \right) }^3} \ .
\eeq
Indeed, no rational identity can express $1 + 7\,t^3 + t^6$ in finite
Euler form so the plethystic logarithm will not terminate.
Now, as promised earlier, we can explain the higher terms such as the 105
and 540. In this example, there are 10 basic invariants and there are
27 relations amongst them. This explains the first 2 terms in
\eref{plogZ3}. This is seen above because if we were to write $1 +
7\,t^3 + t^6$ in Euler form, we would obtain
\beq\label{eulerprodZ3}
1 + 7\,t^3 + t^6 =
\frac{(1-t^6)^{27}(1-t^{12})^{540}\ldots}{(1-t^3)^7(1-t^9)^{105}\ldots}
\eeq
thus we get the $+10t^3$ term from the $(1-t^3)^{3+7}$ factor in the
denominator and the $-27t^6$ term from the $(1-t^6)^{27}$ factor in
the numerator.

Of course, one finds the 27 relations by finding syzygies among
the 10 primitive invariants. The reason we can do this is of course
Theorem \ref{noether} which dictates that we need not go beyond degree
$|G|=3$ to find all basic invariants which generate the entire
invariant polynomial ring. Instead, if we found the syzygies for
the entire invariant ring, we would get the higher terms. That is, we
should, considering the expansion of the Molien series
\beq
\frac{1 + 7\,t^3 + t^6}{{\left(1 - t^3 \right) }^3}
=
1 + 10\,t^3 + 28\,t^6 + 55\,t^9 + 91\,t^{12} +
  136\,t^{15} + 190\,t^{18} + \ldots \ ,
\eeq
consider all $10 + 28 + 55 + \ldots$ invariants as polynomials in 3
variables and find all their syzygies; this should give the higher terms.
Of course this cannot be done all at once, but
nevertheless we can consider the process stepwise: first, syzygies
for $10$ of them, then
$10+28$ of them, etc. In principle, if we
only wish to know up to some degree, we only need to find syzygies for
invariants up to that degree.
This algorithm is the precise analogue of the
infinite product expansion of \eref{eulerprodZ3} into Euler form,
which serves as successive approximation to the LHS in
\eref{eulerprodZ3}.
This also explains the
discrepancies in the case of $\IC^3/\IZ_5$ as seen above. In these
cases where the Euler product is non-terminating, and rational
identities become infinite products, the syzygies should thereby receive
stepwise corrections. We should be able to arrive at the right answer
after some finite number of steps if we only wish to know the terms up
to a desired order.

Let us check up to second order in this example of $\IC^3/\IZ_3$ by
finding the syzygies amongst
the 10 basic invariants of degree 3 and 28 degree 6 invariants. We
find, using \cite{m2}, 595 relations: 55 of degree 6, 225 of degree 9 and
315 of degree 12. This thus reads
\beq
10t^3 + 28 t^6 - 55 t^6 - 225 t^9 - 315 t^{12} =
10t^3 - 27 t^6 - 225 t^9 - 315 t^{12} \ .
\eeq
Good, we reproduce the first 2 terms of $PE^{-1}[M(t)]$ and have the
next 2 terms. This is only up to order 2, i.e., finding syzygies among
$38$ polynomials.
At next order, we would have to find relations among $10+28+55 = 93$
polynomials and correct the $t^9$ and $t^{12}$ coefficients; the
computation becomes increasingly strenuous for the computer\footnote{
This alternative addition of invariants and substraction of relations
is reminiscent of the characters of minimal models and the removal of
null-states using the Kac determinant in 2-dimensional conformal
field theories.}.

%
%------
%
\subsection{Complete Intersections}
We see from the preceeding arguments that the most powerful avatar of
the intimate relations between plethystics and syzygies is realised in
complete intersections, especially in single hypersurfaces. We have
seen that $\Delta(27)$ is one such example of the hypersurface.
The key feature for this
class of varieties is that {\it the series for $f_1 =
  PE^{-1}[f_\infty(t)]$
terminates and is
  polynomial}. This is nice because if we knew the defining equations
and the degrees of the various pieces,
we could re-construct the Poincar\'e series and find the number of
invariants in the gauge theory! This is independent of whether the
variety is orbifold or toric, but should hold in general. 
In fact, we do not even need to know what the gauge theory is!
We shall
see, in \sref{s:mcmahon}, an inverse application of this philosophy,
where we shall construct a variety with desired gauge invariants.

\subsubsection{del Pezzo Family Revisited}\label{s:dP-re}
Take a non-orbifold, non-toric, single hypersurface, the famous cubic
in $\IP^3$; this is the cone over the 6-th del Pezzo surface. From
\eref{f-dpn} and \eref{plog}, we have
\beq
f(t; dP_6) = \frac{1 + t + t^2}{{\left( 1 - t \right) }^3} \Rightarrow
PE^{-1}[f(t; dP_8)] = 4t - t^3 \ ,
\eeq
which says that there should be 4 linear invariants, obeying 1 cubic
relation; precisely the definition of $dP_6$.

Another illustrative example is $dP_8$; here we shall go beyond
projective spaces, but rather to weighted projective spaces. We shall
see that the plethystic logarithm still works. We know
(cf.~\cite{Lerche:1996ni}) that $dP_8$ as a surface is given by a single
equation in $W\IP^3_{1,1,2,3}$. Again, from
\eref{f-dpn} and \eref{plog}, we find
\beq\label{f-dp8}
f(t; dP_8) = \frac{1 - t + t^2}{{\left( 1 - t \right) }^3} \Rightarrow
PE^{-1}[f(t; dP_8)] = 2\,t + t^2 + t^3 - t^6 \ .
\eeq
This is again correct: 2 degree 1, 1 degree 2 and 1 degree 3, obeying
a single degree 6 relation. This can only happen in a weighted
projective space, viz., $W\IP^3_{1,1,2,3}$.
Thus, our proposal \eref{f-dpn} is again confirmed.
We note that, upon comparing \eref{f-dp8} and the result
\eref{del27plog} for $\Delta(27)$, the $f$-functions are the same,
only with the replacement $t \to t^3$. This does not surprise us,
indeed (cf.~e.g.~\cite{Hanany:1999sp,Verlinde:2005jr,Berenstein:2001nk}) 
$\Delta(27)$ is known to be a special point in the
moduli space of $dP_8$'s.

In fact, all del Pezzo surfaces for $n > 4$ are complete intersections
(cf.~e.g., eq.(3.2) of \cite{Lerche:1996ni} and also \cite{hart}). 
We check against
\eref{f-dpn} and find complete agreement. For clarity, let us tabulate
these results:
\beq\label{dp-5678}
\ba{|c|c|c|c|}\hline
dP_n & f = f_\infty(t) & f_1 = PE^{-1}[f(t)] & \mbox{Defining
  Equations} \\ \hline \hline
5 & \frac{1 + 2t + t^2}{{\left( 1 - t \right) }^3} &
   5\,t - 2\,t^2  &  \mbox{2 degree 2 equations in }
   \IP^4 \\ \hline
6 & \frac{1 + t + t^2}{{\left( 1 - t \right) }^3} &
   4\,t - t^3 &  \mbox{1 degree 3 equation in }
   \IP^3 \\ \hline
7 & \frac{1 + t^2}{{\left( 1 - t \right) }^3} &
   3\,t + t^2 - t^4 &  \mbox{1 degree 4 equation in }
   W\IP^3_{1,1,1,2} \\ \hline
8 & \frac{1 - t + t^2}{{\left( 1 - t \right) }^3} &
   2\,t + t^2 + t^3 - t^6 & \mbox{1 degree 6 equation in }
   W\IP^3_{1,1,2,3}
\\ \hline
\ea\eeq
Therefore, for the entire del Pezzo family, members 0 to 3 are checked
by toric methods while 5 to 8 are complete intersections. The only one
remaining is $dP_4$ and from \eref{f-dpn},
\beq
f(t; dP_4) = \frac{1 + 3\,t + t^2}{{\left( 1 - t \right) }^3} =
1 + 6\,t + 16\,t^2 + 31\,t^3 + 51\,t^4 + 76\,t^5 + 106\,t^6 + 141\,t^7
+ 181\,t^8 + 226\,t^9 + \cO(t^{10})
\eeq
predicts the single-trace GIO counting for this variety. The equation
for this variety \cite{Feng:2002fv,hart}
is the (non-complete) intersection of 5 quardrics in $\IP^5$ 
(cf.~also eq. 5.29 of \cite{Butti:2006nk}).
Expanding the plethystic logarithm of $f$ in this case gives
\beq
PE^{-1}[f(t; dP_4)] = 6\,t - 5\,t^2 + 5\,t^3 - 10\,t^4 + 24\,t^5 -
55\,t^6 + 120\,t^7 - \cO(t^8)
\eeq
We see that the first 2 terms are actually correct: there are 5 degree
2 relations in 6 variables! 

For full reference, we tabulate below the other members of the del
Pezzo family, these are non-complete intersections:
\beq\ba{|c|c|c|c|}\hline
dP_n & f = f_\infty(t) & f_1 = PE^{-1}[f(t)] & \mbox{Defining
  Equations} \\ \hline \hline
0 & \frac{1 + 7t + t^2}{{\left( 1 - t \right) }^3} &
  10\,t - 27\,t^2 + 105\,t^3 - 540\,t^4 + 3024\,t^5 + {\cO(t)}^6
     &  (10 | 2^{27})\\ \hline
1 & \frac{1 + 6t + t^2}{{\left( 1 - t \right) }^3} &
  9\,t - 20\,t^2 + 64\,t^3 - 280\,t^4 + 1344\,t^5 + {\cO(t)}^6
     &  (9 | 2^{20}) \\ \hline
2 & \frac{1 + 5t + t^2}{{\left( 1 - t \right) }^3} &
  8\,t - 14\,t^2 + 35\,t^3 - 126\,t^4 + 504\,t^5 + {\cO(t)}^6
     &  (8 | 2^{14})\\ \hline
3 & \frac{1 + 4t + t^2}{{\left( 1 - t \right) }^3} &
  7\,t - 9\,t^2 + 16\,t^3 - 45\,t^4 + 144\,t^5 +  {\cO(t)}^6
     &  (7 | 2^9) \\ \hline
4 & \frac{1 + 3t + t^2}{{\left( 1 - t \right) }^3} &
  6\,t - 5\,t^2 + 5\,t^3 - 10\,t^4 + 24\,t^5 + {\cO(t)}^6
    &   (6 | 2^5) \\ \hline
\ea\eeq
Here, we have computed these defining equations using fat-point
methods on $\IP^2$ \cite{m2,fat}.
We have used the notation, in the above table, that $(m | p_1^{q_1}
\ldots p_k^{q_k})$ denotes $q_1$ equations of degree $p_1$, $q_2$
equations of degree $p_2$, etc., all in $m$ variables. The first
memeber, $dP_0$, is of course $\IC^3/\IZ_3$ as studied in detail in
\eref{c3z3inv} and \eref{c3z3syz}. Furthermore, as mentioned when we
first presented \eref{f-dpn}, $F_0$ has the same $f(t)$ as
$dP_1$. Indeed, we can study the degree 2 Veronese-Segr\`e embedding
of $\IP^1 \times \IP^1$ into $\IP^8$ and see that $F_0$ also has defining
equation in 9 variables as $(9 | 2^{20})$. The precise forms of
these equations, of course, differ from those of $dP_1$. Thus, $dP_1$
and $F_0$ are in different points of a complex structure moduli space.

To compare and contrast, we include the $f_1$ results for the ADE-series
addressed in \sref{s:ADE}; indeed these are all complete intersections
- in fact, single hypersurfaces - so $f_1$, the plethystic logarithm
of the the Molien series should be polynomial:
\beq\ba{|c|c|c|}\hline
G \subset SU(2) & f_1 = PE^{-1}[M(t;~G)] & \mbox{Defining Equation in }
  \IC[u,v,w] \\ \hline \hline
\hat{A}_{n-1} & t^2 + 2t^n - t^{2n} & uv = w^n \\ \hline
\hat{D}_{n+2} & t^4 +t^{2n} + t^{2n+2} - t^{4n+4}
   & u^2 + v^2w = w^{n+1} \\ \hline
\hat{E}_6 & t^6 + t^8 + t^{12} - t^{24} & u^2+v^3+w^{4}=0 \\ \hline
\hat{E}_7 & t^8 + t^{12} + t^{18} - t^{36} & u^2+v^3+vw^3=0 \\ \hline
\hat{E}_8 & t^{12} + t^{20} + t^{30} - t^{60} & u^2+v^3+w^5=0 \\ \hline
\ea\eeq

%%%%%%%%%%%%%%%%%%%%%%%%%%
\subsubsection{Example: The Hypersurface $x^2+y^2+z^2+w^k=0$}\label{s:xyz-w}
%%%%%%%%%%%%%%%%%%%%%%%%%%
Now, let us try another family of complete intersection 3-folds, viz.,
$x^2+y^2+z^2+w^k=0$ in $\IC^4$. For $k=1$, this is just $\IC^3$, 
for $k=2$, it is the conifold $\cC$. 
For $k>2$, the theory is studied in
\cite{Cachazo:2001sg}.
However, for $k \ge 3$, \cite{Gauntlett:2006vf,conti} 
recently showed that there is no
Sasaki-Einstein metric, whereby making the AdS/CFT correspondence a
little ambiguous here.
\comment{
The only cohomogeneity-1 SE 5-folds are Ypq.
}

For $k=2 n$ even,  we have $x,y,z$ being degree $n$ and $w$ being
degree $1$. From this we can read out  $f_1= t+3 t^n- t^{2n}$. Thus we can
calculate that
\[ 
f_\infty(k=2n) = PE[t+3 t^n- t^{2n}]= { (1-t^{2n})\over (1-t)
(1-t^n)^3} \ .
\]
Similarly, for $k$ odd, we have $x,y,z$ being degree $k$ and $w$,
degree $2$. From this we can read out  that 
$f_1= t^2+3 t^k- t^{2k}$ and whence 
\[ 
f_\infty(k)= PE[t^2+3 t^k- t^{2k}]= { (1-t^{2k})\over
(1-t^2) (1-t^k)^3}, \qquad, k \mbox{ odd} \ .
\]
To demonstrate, we list the series expansion of the Poincar\'e series
$f_\infty$ for $k=1$ to $k=5$ and we find
\[
\ba{rcl}
f_\infty(1)& = &  f_\infty(t;~\IC^3) = 
  1+3 t+ 6t^2+10 t^3+1 5 t^4+ 21 t^5+ 28 t^6+ 36 t^7+... \\
f_\infty(2)& = &  f_\infty(t;~\cC) = 
  1+4t+ 9 t^2+ 16 t^3+ 25 t^4+ 36 t^5+49 t^6+ 64 t^7+...\\  
f_\infty(3) & = & 1+ t^2+ 3 t^3+ t^4 + 3 t^5+ 6 t^6 + 3 t^7+ 6 t^8
+ 10 t^9+ 6 t^{10}+... \\  
f_\infty(4) & = & 1+t+4 t^2+ 4 t^3+ 9
t^4+ 9 t^5+ 16 t^6 + 16 t^7 + 25 t^8+ 25 t^9+ 36 t^{10}+36
t^{11}+... \\  
f_\infty(5) & = & 1+t^2+ t^4+ 3 t^5+ t^6+ 3 t^7 +t^8
+3 t^9+6 t^{10}+ 3 t^{11}+6 t^{12}+ 3 t^{13}+ 6 t^{14}+...
\ea\]

\subsection{Refined Relations: The Conifold Revised}
In all of the above, we have used the generating function with a
single variable $t$. How does this all work if everything is
refined fully so as to
contain a tuple of dummy variables for the various $U(1)$-charges?
We shall now see that
the relations are still explicitly encoded by $f_1$. 
We mentioned early on
that the F-term relations are automatically built into the counting.
Indeed, for $\IC^3$, we do not have these
relations - just that $x,y,z$ commute. For the conifold,
we have the simplest demonstration that $f_1$ contains relations. 

Recall the expression for the Poincar\'e series $f_\infty$ in
\eref{f-xyq-coni}. Now, let us take the multi-variate plethystic
logarithm \cite{Labastida:2001ts}, which for $f_\infty(t_1,\ldots,t_m)$ 
is, recollecting \eref{Pnu}
\beq
PE^{-1}[f_\infty(t_1,\ldots,t_m)] =
\sum_{k=1}^\infty  \frac{\mu(k)}{k} \log
(f_\infty(t_1^k,\ldots,t_m^k)) \ .
\eeq
The result is
\beq\label{f1-xyq-coni}
f_1 = PE^{-1}[{xy (1-q^2)\over (1-q x)(1-q y) (q-x)(q-y)}] = 
\frac{q}{x} + q\,x + \frac{q}{y} + q\,y -q^2 \ .
\eeq

Indeed, $f_1$ is polynomial because the conifold is complete
intersection. There are four invariants, corresponding to 
$q x, \frac{q}{x}, q y,  \frac{q}{y}$. 
If we would write just these generators without any subtractions, this
would be merely the result for $\IC^4$.
Therefore it is not enough. To put the relation, we notice that
$(q x) (q/x) = (q y) (q/y) = q^2$ and therefore
we should subtract one combination of $q^2$. We thus reproduce
\eref{f1-xyq-coni}.
%This is a simple procedure and for the conifold it is enough to
%generate all multi-trace operators for $N=1$, i.e., to give $g_1$.
The procedure is
simple and analogous for complete intersections. However, for
non-complete intersections, once we
make the subtraction we are taking away too much. We must therefore
compensate by adding those which are subtracted, etc. ad infinitum,
just like the non-terminating series explained in \sref{s:syn}.

%
%========= ASYMPTOTICS
%
\section{Asymptotics and the Meinardus Theorem}\label{s:asym}
\setall
We have encountered, in the preceeding discussions, many infinite
products of Euler type. Indeed, we recall that the
generating function for multi-trace GIO's is
\beq\label{pn-expand}
g(t) := \sum\limits_{n=0}^\infty p_n t^n
= PE[f(t)] = \frac{1}{\prod\limits_{n=1}^\infty (1-t^n)^{a_n}}
\qquad \mbox{ for } \qquad
f(t) = \sum\limits_{m=0}^\infty a_m t^m \ .
\eeq
In the case that all $a_n = 1$, $g(t)$ is the Euler function, or, up
to a factor of $t^{-\frac{1}{24}}$, the Dedekind $\eta$-function.
This is our familiar generating function for the number of ways of
partitioning integers. The Hardy-Ramanujan equation gives the
asymptotic behaviour of $p_n$ and was what gave rise to the
Hagedorn temperature (q.v.~\cite{GSW}).
It is, needless to say, important to find analogous asymptotic behaviours
for general $a_n$. This would give micro-state counting for our quiver
gauge theories.

Luckily, this generalisation of Hardy-Ramanujan is known.
This is a result due to G.~Meinardus \cite{meinardus}
(q.v.~\cite{actor}, to whose notation we adhere,
for some explicit results). Meinardus' theorem states that the
asymptotic behaviour of $p_n$ in \eref{pn-expand} is:
\beq\label{mein}
p_n \sim C_1 n^{C_2} \exp
\left[ n^{\frac{\alpha}{\alpha+1}}(1+\frac{1}{\alpha})\left(A
\Gamma(\alpha+1) \zeta(\alpha+1)\right)^{\frac{1}{\alpha+1}}
\right] (1 + \cO(n^{-C_3})) \ ,
\eeq
if the Dirichlet series for the coefficients $a_m$ of $f$,
defined as
\beq
D(s) := \sum\limits_{m=1}^\infty \frac{a_m}{m^s}, \qquad
{\rm Re}(s) > \alpha > 0 \ ,
\eeq
converges and is analytically continuable into the strip $-C_0 <
\mbox{Re}(s) \le \alpha$ for some real constant $0 < C_0 < 1$ and such
that in this strip, $D(s)$ has only 1 simple pole at $s = \alpha \in
\IR_+$ with residue $A$. The constants in \eref{mein} are
\bea\nn
C_1 &=& e^{D'(0)} \frac{1}{\sqrt{2\pi(\alpha+1)}} \left(
    A \Gamma(\alpha+1)
    \zeta(\alpha+1)\right)^{\frac{1-2D(0)}{2(\alpha+1)}}, \\
C_2 &=& \frac{D(0)-1-\frac{\alpha}{2}}{\alpha+1} \ ,
\eea
and $C_3$ some positive constant.

\subsection{Example: $\IC$ and Dedekind $\eta$}\label{s:eta}
For example, when all $a_m = 1$, we have the usual partition of
integers and the Dirichlet series is just the Riemann
$\zeta$-function. The generating function $f(t) =
\sum\limits_{m=0}^\infty t^m$ is of course simply $\frac{1}{1-t}$ and
the geometry is that of the complex line $\IC$.
Using the above results of Meinardus, we have
\[
\alpha=1, \quad A = 1, \quad
D(0) = -\frac12, \quad D'(0) = e^{-\frac12 \log(2\pi)}; \qquad
C_1 = \frac{1}{4\sqrt{3}}, \quad C_2 = -1 \ ,
\]
giving us
\[
p_n \sim \frac{1}{4\sqrt{3} n} e^{\pi \sqrt{\frac{2n}{3}}} (1 +
\cO(n^{-C})) \ ,
\]
precisely the Hardy-Ramanujan result.

\subsection{Example: MacMahon Function and a Riemann Surface}
\label{s:mcmahon}
Next, consider
\[
f(t) = \frac{1-t+t^2}{(1-t)^2},~a_n=n, \quad \Rightarrow \quad g(t) =
\frac{1}{\prod\limits_{n=1}^\infty (1-t^n)^{n}} \ .
\]
As $PE^{-1}[f(t)]=t+t^2+t^3-t^6$, this is a complete intersection,
given as a hypersurface of degree 6 in $W\IP^2_{1,2,3}$. The dimension
is therefore 1 and is hence a Riemann surface. The total space is an
affine cone over this surface and is of dimension 2.
\comment{
Using the degree-genus formula, $g = \frac12(d-1)(d-2)$, the genus of
the Riemann surface is 10.
}
We can embed $W\IP^2_{1,2,3}$, through a Veronese-Segr\'e map, into
$\IP^6$ and see that the genus of the Riemann surface is
1. Alternatively, we can projectivise the weight-one coordinate in
$W\IP^2_{1,2,3}$ and simply obtain a standard elliptic
curve. Therefore, the geometry is a cone over a torus!
The generating function $g(t)$ is the well-known MacMahon function
\cite{Okounkov:2003sp}. We see that
\[
\alpha=2, \quad A = 1, \quad
D(0) = -\frac{1}{12}, \quad D'(0) = \frac{1}{12}-\log(G_l); \qquad
C_1 = \frac{e^{\frac{1}{12}}\,
    {\zeta(3)}^{\frac{7}{36}}}{2^
     {\frac{11}{36}}\,{G_l}\,
    {\sqrt{3\,\pi }}}, 
\quad C_2 = -\frac{25}{36} \ ,
\]
where $G_l := e^{\frac{1}{12} - \zeta'(-1)}
\simeq 1.28243$ is the Glaisher constant.
Hence,
\beq
p_n \sim \frac{e^{\frac{1}{12}}\,
    {\zeta(3)}^{\frac{7}{36}}}{2^{\frac{11}{36}}\,G_l\,
    {\sqrt{3\,\pi }}} n^{- \frac{25}{36}}
\exp\left( \frac{3\,{\zeta(3)}^{\frac{1}{3}}}
  {2^{\frac{2}{3}}} n^{\frac23}\right) \ .
\eeq

\subsection{Example: Our Familiar $\IC^3$ }
Returning to something we have encountered earlier,
let us attack the $\IC^3$ example of \eref{g-C3}. We can now
find the coefficients $d_k$ therein! Using \eref{eulerPE} we have that
\beq
g(t;~\IC^3) = PE[f(t;~\IC^3)] =  \frac{1}{\prod\limits_{n=1}^\infty
  (1-t^n)^{a_n}}, \qquad a_n = \frac12(n+1)(n+2) \ .
\eeq
We can readily see that $D(s) = \frac12\left(\zeta(-2 + s) +
3 \zeta(-1 + s) + 2\zeta(s)\right)$. We see that there are 3 poles, at
1,2 and 3. Of course, Meinardus Theorem requires that there be only
one pole within a strip. Thus, one must consider one monomial of $a_n$
at a time and consider the break-down
\[
g(t;~\IC^3) =
\frac{1}{\prod\limits_{n=1}^\infty(1-t^n)^{a_{n_1}}} \cdot
\frac{1}{\prod\limits_{n=1}^\infty(1-t^n)^{a_{n_2}}} \cdot
\frac{1}{\prod\limits_{n=1}^\infty(1-t^n)^{a_{n_3}}}
:=g_1(t)g_2(t)g_3(t),
\]
with
$a_{n_1} = \frac12n^2,~a_{n_2} = \frac32n,~a_{n_3} = 1$.
Applying Meinardus and defining
$g_{i=1,2,3} := \sum\limits_{n=0}^\infty p_i(n) t^n$, we have that
\beq\ba{l}
p_1(n) \sim \frac{e^{\frac{\zeta'(-2)}{2}}}
   {2\,2^{\frac{5}{8}}\,{15}^{\frac{1}{8}}}
   n^{-\frac58} \exp\left( \frac{2\,2^{\frac{3}{4}}\,\pi }
   {3\,{15}^{\frac{1}{4}}} n^{\frac34}\right), \quad
p_2(n) \sim \frac{e^{\frac{1}{8}}\,
     {\zeta(3)}^{\frac{5}{24}}}{3^
      {\frac{7}{24}}\,{G_l}^
      {\frac{3}{2}}\,{\sqrt{2\,\pi }}}
   n^{-\frac{17}{24}} \exp\left( \frac{3\,{\left(
       3\,\zeta(3) \right) }^{\frac{1}{3}}}{2} n^{\frac23}\right),
   \\
p_3(n) \sim \frac{1}{4 \cdot 3^{\frac38}}
   n^{-\frac78} \exp\left(
       \frac{\pi}{\sqrt{3}}n^{\frac12}\right) \ .
\ea\eeq
Therefore, we have the convolution $p(n) = \sum\limits_{r+s+t=n}
p_1(r) p_2(s) p_3(t)$ and  
since the exponential growth of $p_1(n)$ dominates over the other
two, for large $n$
\beq
p(n) \sim p_1(n) \ .
\eeq

%%
%
%  FINITE N
%%%
\section{Single-Trace and Multi-Trace for Finite $N$}\label{s:finiteN}
\setall
We have, in all preceeding discussions, made the important
simplification of taking $N$, the matrix size of the operators, to
infinity, whereby decoupling spurious relations which arise from the
lack of commutativity among the matrices of finite size. As mentioned in
the introduction, the problem of counting for finite $N$ is a
significantly more difficult one. Nevertheless, we shall see in this
section that the plethystics are still applicable.

\newcommand{\ag}{\alpha}
\newcommand{\bg}{\beta}
\newcommand{\cg}{\gamma}
\newcommand{\dg}{\delta}
\newcommand{\agp}{\dot{\ag}}
\newcommand{\bgp}{\dot{\bg}}
\newcommand{\cgp}{\dot{\cg}}
\newcommand{\dgp}{\dot{\dg}}

We consider the problem of counting BPS states of $\cN=1$ supersymmetric quiver
gauge theories for $N$ finite; $N$ is the number of D3-branes at the
tip of the CY cone. We denote the generating function for
multi-trace GIO's by $g_N$. This problem is of significant interest,
for instance, to studying phase transitions and $AdS_5$ black
holes. We already considered in the previous sections the functions
$g_1 = f_\infty$ and $g_{\infty} = PE[g_1]$, and we are going to
%show
propose
that it is still quite simple to reconstruct $g_N$ in terms of
$g_1$.

Suppose the single-trace generating function is given by $g_1(t) =
f_\infty(t) = \sum\limits_{n=0}^{\infty} a_n t^n$, then we can
construct the following function:
\beq\label{Zdef}
g(\nu ; t) := \prod_{m=0}^{\infty} \frac{1}{(1 - \nu  \, t^m)^{a_m}} \ .
\eeq
We immediately notice a strong similarity to the Euler product form of
the plethystic exponential introduced in \eref{pexp}, \eref{eulerPE}
and especially \eref{Pnu}. 
\comment{
The only difference between \eref{Pnu} and
\eref{Zdef}
is that in the former the sum is from $1$ to $\infty$ while for the
latter it is from $0$ to $\infty$.
}

We now propose that the finite $N$ multi-trace generating
function $g_N(t)$ is simply given by the expansion
\beq\label{ZNdef}
\sum_{N=0}^\infty g_N(t) \nu^N = g(\nu ; t)  \ .
\eeq
We have 2 limiting cases to check, viz., $g_1$ and $g_\infty$, with
which we are now quite familiar.
First, we note that
\[
\partial_\nu g(\nu ; t) =
 \sum\limits_{k=0}^\infty  \frac{(-a_k)(-t^k)}{(1 - \nu  \, t^k)^{a_k+1}}
 \prod_{m = 0; m\neq  k}^\infty  \frac{1}{(1 -  \nu \, t^m)^{a_m}}=
g(\nu ; t)
 \sum_{k=0}^\infty  \frac{a_k t^k}{(1 - \nu \, t^k)} \ .
\]
Furthermore, since $g(0 ; t) = 1$, we have that 
$\partial_\nu g(0 ; t) = \sum_{k=0}^\infty a_k t^k$.
Therefore, the coefficient of $\nu$ in \eref{ZNdef} is indeed our 
$g_1$:
\beq\label{fdZp} 
\partial_\nu g(\nu,q) |_{\nu = 0} = g_1(q) \ . 
\eeq

Next, let us check whether the $N$-th coefficient for $N \to \infty$
gives our $g_\infty$. This coefficient can be found by considering the
limit\footnote{In this paper we always have
  $a_0=1$, since the only operator of vanishing scaling dimension is
  the identity. We can see this explicitly in all the examples we have
  given throughout.}  
$\lim\limits_{\nu\rightarrow 1} (1-\nu)^{a_0} g(\nu ; t)$ which
extracts the large $N$-term in the series expansion. We see that
\beq\label{PErelation}
 \lim_{\nu\rightarrow 1} (1-\nu)^{a_0} g(\nu ; t) = \prod_{m=1}^{\infty}
 \frac{1}{(1 - \, t^m)^{a_m}} \equiv  PE[g_1(t)] \ . 
\eeq
Therefore, our expansion \eref{ZNdef} has the property that its large
$N$ coefficient is the $PE$ of the linear coefficient, precisely what
is required of $g_\infty$.
We will see in the ensuing text why the expansion does what it is
supposed to.
\comment{
We now confess that we do not yet have a rigorous derivation 
that the expansion \eref{ZNdef} indeed gives $g_N$ for all $N$, but
from the following examples it can be seen that it works very well. It
is indeed an interesting problem in combinatorics to have the full
proof.}

It is very interesting to compare \eref{Pnu} and \eref{Zdef}.
From it we can see that the parameter $\nu$ in \eref{Zdef} can have two
different interpretations:
\begin{itemize}
\item[I.] It counts the number of single-trace GIO's in a
 multi-trace GIO for the limit of matrix rank $N\to \infty$ (here we
 include the 
single trace of identity as well) as in \eref{Pnu};
\item[II.] It counts the number of multi-trace GIO's for matrix rank
$N$ given by the finite power of $\nu$ as in \eref{ZNdef}.
\end{itemize}
Naively these two counting problems seem to be unrelated, but
our proposed formula \eref{Zdef} indicates that they are the same.
\comment{It will be very interesting to understand their relation more
concretely and to find a proof for proposal \eref{Zdef} and
\eref{ZNdef}.}

Formulae (\ref{Zdef}) and (\ref{ZNdef}) give the
general solution for counting multi-trace BPS GIO's, for a finite
number $N$ of D3-branes. In fact, the relation between $f_N$ and
$g_N$, in general, still obeys the plethytic exponential as was in
\eref{g1-f1} and \eref{pexp}, which we summarise now  (for a list of
variables $t_i$):
\beq\label{fN-gN}
g_N(t_i) = PE[f_N(t_i)] =
\exp\left( \sum_{k=1}^\infty \frac{f_N(t_i^k) - f_N(0,\ldots,0)}{k}
\right) \ .
\eeq

%============
\paragraph{Symmetric Products and Moduli Spaces: }
We can in fact re-examine the finite $N$ counting from another
perspective.
The standard lore for $N$ D3-branes probing a CY manifold, $X$, is
that the moduli space of vacua, ${\cal M}_{\rm vac}(N;X)$, is the
symmetric product of $N$ copies of the CY manifold,
\beq\label{symN-X}
{\cal M}_{\rm vac}(N;X) = {\cal S}^N(X) := \frac{X^N}{S_N},
\eeq
where $S_N$ is the permutation group of $N$ elements.
Following our general line in this paper we can now state two
important relations:
\begin{enumerate}
\item $g_N$ counts multi trace operators for one D3-brane on the
  symmetric product of $N$ CY manifolds ${\cal M}_{\rm vac}(N;X)$:
\beq
g_N(t;X) = g_1 (t ; \frac{X^N}{S_N}) = f_\infty (t ; \frac{X^N}{S_N}).
\label{multitracesym}
\eeq
Alternatively we can think of it as the Poincar\'e series for ${\cal
  M}_{\rm vac}(N;X)$. Furthermore, from the second equality we
conclude that $g_N$ also counts the single trace operators on ${\cal
  M}_{\rm vac}(N;X)$ in the limit in which there are no matrix
relations at all, $N\rightarrow\infty$. 

\item $f_N$ counts single trace operators for one D3-brane on the
  symmetric product of $N$ CY manifolds ${\cal M}_{\rm vac}(N;X)$: 
\beq
f_N(t;X) = f_1 (t ; \frac{X^N}{S_N}).
\label{singletracesym}
\eeq
Here, we are again using the plethystic exponential relations.
In fact, in cases in which the symmetric product is a complete
intersection, $f_N$ will be finite and we can compare the computation
of $f_N$ using the formulas at the beginning of this section to
independent derivations using the property that the manifold is a
complete intersection.
In cases in which the symmetric product is not a complete intersection
we can still use the reasoning of \sref{s:syn}.
To count the number of generators and the number of defining relations
for the symmetric product.
\end{enumerate}

Actually, \eref{multitracesym} is the reason why \eref{ZNdef} works. The
far-LHS of the expression is the generating function for multi-trace
at finite $N$, in line with interpretation II stated above, 
while the far-RHS is the
single-trace generating function at $N \to \infty$, in accord with
interpretation I. Indeed, multi-trace operators with $N$ single-trace
components in $X$ is in one-to-one correspondence with single-trace
operators in $Sym^N(X)$. Therefore, \eref{multitracesym} serves to bridge the
two, whereby showing that the expansion coefficients $g_N$ indeed
count multi-trace operators at finite $N$.
Having delved into much abstraction, let
us be concrete and
now show how these proposals agree with known results.

%%----------
\subsection{Example: The Complex Line $\IC$}
The simplest example, as was encountered in \sref{s:eta}, is given
by
\beq
g_1(t) = f_\infty(t;~\IC) = \frac{1}{(1-t)} = \sum_{n=0}^{\infty}t^n
\ . 
\eeq
This is the well known partition function of the half-BPS
states in $\cN=4$ SYM (given a particular choice of the
supercharges). This partition function also counts the ``extremal''
BPS mesons in toric quivers (i.e. the mesons lying along an edge of
the toric cone). In this case, it should be simple to check the
multi-trace generating function is, as dictated by \eref{Zdef} and
\eref{ZNdef}, given precisely by 
\beq
g(\nu ; t) = \sum_{N=0}^{\infty} g_N(t) \nu^N = \prod_{m=0}^{\infty}
\frac{1}{(1 - \nu t^m)} \ . 
\eeq
We note that $g_N$ is also the partition
function of $N$ bosonic one-dimensional harmonic oscillators.
In other words, we are taking a quantum particle whose single
particle states are precisely the integer points in the
half-line $\IZ_{\geq 0}$, and considering the placement of $N$ of such
bosons.  We can obtain $g_N$ for any value of $N$ by Taylor expansion:
\beq\label{gN-C}
g_N(t) = \prod_{n=1}^N\frac{1}{1-t^n}.
\eeq

In fact, there is another way to see \eref{gN-C}.
Indeed, we have the single-trace generating function explicitly:
\beq
f_N(t;~\IC) = 1 + t + t^2 + ... + t^N = \frac{1 - t^{N+1}}{1-t}\ ,
\label{foneC}
\eeq
which encode the operators $\tr(X^i)$ for $i=0,\ldots,N$. We can take
the PE of \eref{foneC} and using the multiplicative property
\eref{eulerPE}, arrive at \eref{gN-C} directly.

A few specific cases are at hand. For $N=1$,
\beq
g_1(t) = \frac{1}{1-t} =  1 + t + t^2 + \ldots + t^n + \ldots\;,
\eeq
corresponding to the operators
\[
\II;\; \tr(X);\; \tr(X)^2;\; \ldots ;\; \tr(X)^n;\; \ldots \ . 
\]
For $N=2$ we get \beq
g_2(t) = \frac{1}{(1-t)(1-t^2)} = 1 + t + 2 t^2 + 2 t^3 + 3 t^4 + 3
t^5 + 4 t^6 + \ldots + (n+1) t^{2n} + (n+1) t^{2n+1} +  \ldots\;,
\eeq
corresponding to the operators (we have dropped the $\tr$ in the
notation without ambiguity): 
\[\ba{l} 
\II;\; \qquad (X);\; \qquad
(X)^2,(X^2);\; \qquad (X)^3,(X)(X^2);\;  \qquad
(X)^4,(X)^2(X^2),(X^2)^2;\; \ldots\\  \ldots;\;
(X)^{2n},(X^2)(X)^{2n-2},(X^2)^2(X)^{2n-4},\ldots,(X^2)^{n-1};
\ldots  \ . 
\ea\]
Indeed $g_2(t)$ is the Poincar\'e series for $\IC^2/\IZ_2$ where
$\IZ_2$ acts by the exchange of the two coordinates $(z_1,z_2)$ of
$\IC^2$. It has two generators, one of degree 1 corresponding to
$a := z_1+z_2$ and another of degree 2, corresponding to
$b := z_1z_2$. It is easy to see that the other invariant of degree 2 is
represented in terms of these two, $z_1^2+z_2^2 = a^2-2b$. Similarly,
all other invariants of higher degree can be written in terms of these
two.

For $N=3$, we have 
\beq
g_3(t) = \frac{1}{(1-t)(1-t^2)(1-t^3)} = 1 + t + 2 t^2 + 3 t^3 + 4 t^4
+ 5 t^5 + 7 t^6 + 8 t^7 + 10 t^8 + 12 t^9 + 14 t^{10} +
\cO(t^{11})\;, 
\eeq
corresponding to the operators 
\[\ba{l} \II; \qquad (X); \qquad (X^2),
(X)^2; \qquad 
(X^3), (X)(X^2), (X)^3; \qquad  (X)(X^3), (X^2)^2, (X)^2(X^2),(X)^4; \\
(X^2)(X^3), (X)^2(X^3), (X)(X^2)^2, (X)^3(X^2),(X)^5; \ldots \ . 
\ea\]
We see that indeed our generating function \eref{ZNdef} agrees with the
explicit counting.

To demonstrate the interplay between plethystics and symmetric
products, we now calculate $f_N$ using \eref{singletracesym}. We need
to find $f_N(t;~\IC) = f_1 (t ; \frac{\IC^N}{S_N})$.
Now, if we expand an $N$-th order polynomial equation in one
complex variable $x$,
\beq
P_N(x) = x^N + \sum_{i=1}^{N}a_ix^{N-i} = \prod_{j=1}^N(x-z_j),
\eeq
we find that the parameters $a_i$, $i=1\ldots N$ are symmetric
functions of degree $i$ for the coordinates $z_j$ on $\IC^N$.
That is, $a_i$ are coordinates on ${\cal M}_{\rm vac}(N; \IC) =
\IC^N/S_N$. Furthermore, there is precisely one generator of degree $i$
for the ring of symmetric functions of the $z_j$ for any $i$ between 0
and $N$. We can pick the generators to be the coordinates $a_i$. Any
other symmetric function of degree $i>N$ can be written in terms of
the $a_i$. Collecting this together we find $f_1$ for ${\cal M}_{\rm
  vac}(N ; \IC)$ as in \eref{foneC},
\beq
f_1 (t ; \frac{\IC^N}{S_N}) = \frac{1-t^{N+1}}{1-t} = f_N(t;\IC),
\eeq
consistent with our proposal from \eref{singletracesym} which implies,
using the plethystic exponential, \eref{multitracesym}, thus
supporting the proposal for multi-trace.

%===
\subsection{Example: The Complex Plane $\IC^2$}
Next, we address a slightly more involved example, viz., $\IC^2$.
This case is quite simple as well and describes $1/4$-BPS operators
in $U(N)$ $\cN=4$ SYM. This also describes a subsector of BPS
operators in many toric quivers, namely the operators corresponding
to points lying along a face of the toric cone (this face gives a
toric subcone of the toric cone, a cone over a SUSY $3$ cycle), when
the SUSY 3-cycle has the topology of $S^3$. Now, we have that
\bea
f_1(t;~\IC^2)&=&2t; \cr
g_1(t;~\IC^2) &=& PE[f_1(t;~\IC^2)] = \frac{1}{(1-t)^2} = \sum_{n=0}^{\infty} (n+1) t^n \ .
\eea
Formulae (\ref{Pnu}) and (\ref{Zdef}) takes the form
\beq
g(\nu ; t) = \prod_{m=0}^\infty\frac{1}{(1-\nu t^m)^{m+1}}
= \exp(\sum_{k=0}^\infty \frac{\nu^k}{ k (1-t^{2k})})
\eeq
which looks deceptively similar to the generalized MacMahon function
which is used as the partition function for the topological string on
the conifold  \cite{Halmagyi:2005vk}.
Using \eref{ZNdef}, we get
\beq
g_2(t) = \frac{1+t^2}{(1-t)^4(1+t)^2} = 1 + 2 t + 6 t^2 + 10 t^3 + 19
t^4 + 28 t^5 + 44 t^6 +  \ldots
\ . \eeq
We report the $R$-charge $3$ GIO's, corresponding to the term
$10t^3$. We see that indeed there are $10 = 2 + 3 + 3 + 2$ of them: 
\[
(X^2)(X), (X)^3;\qquad  (X^2)(Y), (X)(X Y), (X)^2 (Y);\qquad \left[ X
  \leftrightarrow Y \right] \ . 
\]
Next, for $R$-charge $4$ GIO's, we see
that there are indeed $19 = 3 + 4 + 5 + 4 + 3$ them: \begin{eqnarray*}
(X^2)^2, (X^2) (X)^2, (X)^4; \phantom{aaaaaaaaaaaa}\\
(X^2) (X) (Y), \quad (X)^2 (X Y), \quad (X^2) (X Y), \quad (X)^3 (Y);
  \phantom{aaaa}\\
(X^2) (Y^2), \quad (X^2) (Y)^2, \quad (X)^2 (Y^2), \quad
  (X) (X Y) (Y), \quad (X)^2 (Y)^2;\\
 \left[ X \leftrightarrow Y \right] \phantom{aaaaaaaaaaaaaaaaaaa}\\
\end{eqnarray*}
The moduli space of vacua for this case is $(\IC^2)^2/\IZ_2$ where the
$\IZ_2$ acts as exchange of the two coordinates. It is a complete
intersection and has 
\beq
f_1 (t ; (\IC^2)^2/\IZ_2) = f_2 (t ; \IC^2) = 1+2t +3t^2-t^4.
\eeq

%%%%%%%%%%%%%%%%%%%%%%%%%%%%%%%
\subsection{Example: The Conifold}
%%%%%%%%%%%%%%%%%%%%%%%%%%%%%%%%
The single trace GIO's for the conifold are given, recalling
\eref{P-coni}, by
\beq\label{z1con}
g_1(t) = \frac{1+t}{(1-t)^3} = \sum_{n=0}^{\infty} (n+1)^2 t^n \ .
\eeq 
From formula (\ref{Zdef}) we get 
\beq\label{z2con}
f_2(t) = 
\frac{1 + t + 7\,t^2 + 3\,t^3 + 4\,t^4}
  {{\left( 1 - t \right) }^3\,
    {\left( 1 - t^2 \right) }^3}
=
1 + 4 t + 19 t^2 + 52 t^3 + 134 t^4 + 280 t^5 + 554 t^6 +
\ldots\;,
\eeq
corresponding to the operators (again, we drop the ${\rm Tr}$ for
brevity):
\beq
\II; \qquad
(M_{i,j});  \ldots \ . \eeq
At $R$-charge $2$ we have $9$ single-trace
GIO's (cf.~\fref{f:conidimer}) and $10$ double-trace GIO's, given
explicitly by:
\beq\label{conifops-1} \ba{ccc}
(M_{0,1}M_{0,1}) & (M_{0,1}M_{1,0}) &
(M_{1,0}M_{1,0})\\ (M_{0,1}M_{-1,0}) & (M_{0,1}M_{0,-1})=
~(M_{1,0}M_{-1,0}) & (M_{1,0}M_{0,-1}) \\
(M_{-1,0}M_{-1,0}) & (M_{-1,0}M_{0,-1}) & (M_{0,-1}M_{0,-1})
\ .
\ea\eeq
and
\beq\label{conifops-2}\ba{ccc}
(M_{0,1})(M_{0,1}) & (M_{0,1})(M_{1,0}) &
(M_{1,0})(M_{1,0})\\ (M_{0,1})(M_{-1,0}) & (M_{0,1})(M_{0,-1}),
~(M_{1,0})(M_{-1,0}) & (M_{1,0})(M_{0,-1}) \\
(M_{-1,0})(M_{-1,0}) & (M_{-1,0})(M_{0,-1}) & (M_{0,-1})(M_{0,-1})
\ .
\ea\eeq
We emphasize that all these GIO's have vanishing mesonic charge, we
are not counting the BPS operators such as $\det(A)$.

%%%%%%%%%%%%%%%
\subsection{Refinement: Multicharges at Finite $N$}
%%%%%%%%%%%%%%%%%

As with the refinement of the charges discussed in\sref{s:refine},
it is simple to generalize the arguments of the previous subsection
to partition functions $g_1$ depending on more than one variables,
arising for instance from CY cones with isometry $U(1)^2$ or
$U(1)^3$.

Consider a toric CY cone whose integer points are described by the
set $C$. The single particle states are described by the generating
function 
\beq  g_1(t_1,t_2,t_3)= \sum\limits_{n,m,r \in C} t_1^n
t_2^m t_3^r \ . 
\eeq
Every point in $C$ contributes once to $g_1$,
i.e., we are considering a quantum particle whose states are
precisely the integer points in the toric cone. The multi-trace
generating function $g_N(t_1,t_2,t_3)$, in analogy with \eref{Zdef},
is given by 
\beq\label{toricmt}
g(\nu,t_i) = \sum_N g_N(t_i) \nu^N = \prod_{n,m,r \in C}
\frac{1}{(1 - \nu  t_1^n  t_2^m  t_3^r)} 
= \exp \left(\sum_{k=1}^\infty \frac{\nu^k}{k} g_1(t_1^k,t_2^k,t_3^k) \right)
\ . 
\eeq
The coefficients $g_N(t_i)$ can be interpreted as the multi-particle
partition function of $N$ boson whose single particle states are given
by the integer points of $C$.

%----
\subsubsection{The Conifold Reloaded}
Recalling \eref{Pxyz-coni}, the generating function $g_1$ is given by:
\[
\hspace{-1cm}
g_1 =  {xy (1-q^2)\over (1-q x)(1-q y) (q-x)(q-y)} = 1+
q(x+y+{1\over x}+{1\over y})+ q^2 ( 1+{1\over x^2}+x^2
+{1\over y^2}+y^2 +x y+ {1\over x y}+{x\over y}+{y\over x})+ \ldots \ .
\]
We can identify the charges $(q,x,y)$ for following four operators
\[ M_{0,1}= (1,
1,0),~~~M_{0,-1}=(1,-1,0),~~~M_{0,1}=(1,0,1),~~~M_{0,-1}=(1,0,-1) \ .
\]
Therefore, the generating function is, according to \eref{toricmt},
\bean 
g(\nu; q,x,y)& = & {1\over (1-\nu)} {1\over (1-\nu q x)(1-\nu q y)(1-\nu
{q\over x})(1-\nu {q\over y})}{1\over (1-\nu q^2)(1- \nu q^2 x^2)(1-\nu
{q^2\over x^2})} \\ & & {1\over (1- \nu q^2 y^2)(1-\nu {q^2\over
y^2})(1-\nu q^2 x y)(1-\nu {q^2\over x y})(1-\nu {q^2 x\over y})(1-\nu {q^2
y\over x})}  \ .
\eean

Now we try to apply above result. For $N=1$ case we find the
\[
g_1=1 + q\,\left( \frac{1}{x} + x + \frac{1}{y} + y \right)  +
 q^2\,\left( 1 + \frac{1}{x^{2}} + x^2 + \frac{1}{y^{2}} + \frac{1}{x\,y}
 + \frac{x}{y} + \frac{y}{x} + x\,y + y^2 \right)+... \ ,
\]
which has obvious correspondence with the variables $M_{i,j}$.
For $N=2$, we get the following expansion up to $R$-charge two
\[
g_2=1 + q\,\left( \frac{1}{x} + x + \frac{1}{y} + y \right)  +
 q^2\,\left( 3 + \frac{2}{x^2} + 2\,x^2 + \frac{2}{y^2} +
\frac{2}{x\,y} + \frac{2\,x}{y} + \frac{2\,y}{x} +
    2\,x\,y + 2\,y^2 \right) \ .
\]
Again, it is easy to find the mapping between the terms here and
operators in \eref{conifops-1} and \eref{conifops-2}.

%%-----------------
\subsection{Theories with only $U(1)^2$ symmetry}
We can also start from a theory with only $U(1)^2$ symmetry, whose
Poincar\'e series is given by 
\beq  g_1(t_1,t_2)= \sum_{n,m \geq 0}
a_{m,n} t_1^n t_2^m \ .
\eeq
Using this we can find the generating function given by
\beq
g(\nu,t_i) = \sum_N g_N(t_i) \nu^N =
\prod_{n,m \geq 0}  \frac{1}{(1 - \nu t_1^n  t_2^m)^{a_{m,n}}} =
\exp \left(\sum_{k=0}^\infty \frac{\nu^k}{k} g_1(t_1^k,t_2^k) \right)
\ .
\eeq

{\vskip 1in}

%==========
% CONCLUSIONS
%==========
\section{Conclusions and Prospects}\label{s:conc}\setall
In this paper we have considered the 1/2-BPS operators of generic
 superconformal quiver gauge theories, living on $N$ D3-branes probing
 the tip of a Calabi-Yau (CY) cone. It is shown how to construct the
 explicit generating functions that count the scalar BPS operators. We
 have discussed in great detail various classes of CYs (orbifolds, toric 
 varietes, del Pezzo's and complete intersections, even geometries for
 which the gauge theory is not yet known), providing a 
 simple bridge (the ``Plethystic Exponential'') between the algebraic 
 geometry of the CY and the generating functions of the BPS states.
 
The plethystics directly relate three different 
generating functions: (1) the defining equations of the CY 
(syzygies) as well as the moduli space of vacua, 
(2) the single-trace operators and 
(3) the multi-trace 
operators. Beautiful structures thus emerge, exhibiting a rich
inter-play between quiver gauge theories, algebraic geometry,
combinatorics and analytic number theory.
This intricate framework allows us to solve the 3 problems posed in
the introduction, whereby realising our wish-list.

There are a number of directions that could be pursued for future
work. Let us discuss some of them.
We only considered the subset of operators with vanishing baryonic
charges. For instance, for the
conifold we did not include, in the counting, the operator $\det(A)$. 
It would be nice to find the partition functions including the
baryonic charge, that may 
be compared to analogous computations on the string, $AdS_5 \times
X^5$, side.

A possible continuation of our work could be to extend the study of
chiral 1/2-BPS operators in $\cN=1$ quivers
to consider also 1/2-BPS operators with space-time angular momenta and
1/2-BPS fermionic operators.
This would give partition functions depending on additional charges
and would, for instance, enable a computation of the BPS index in
quiver gauge theories, see
\cite{Romelsberger:2005eg,Kinney:2005ej,Nakayama:2005mf,Biswas:2006tj}.

Another exension would be to consider 1/4-BPS operators,
annihilated only by one supercharge. We remark that we are studying 
1/2-BPS operators in $\cN=1$ gauge theories annihilated by 2 out of
the 4 supercharges $Q$.
These are the analogues of 1/8-BPS operators (annihilated by 2 out of
the 16 supercharges) in $\cN=4$ SYM. It would be very interesting to
extend the study to 1/4-BPS operators of quivers (annihilated by
only 1 supercharge $Q$, analogous to 1/16-BPS ops in $\cN=4$ SYM)
\footnote{One single-trace example of such operators is given by 
  $\tr(O\,K)$, where $O$ is a scalar BPS operator and $K$ is the scalar
  SUSY partner of a conserved current:
  $\bar{Q}_{\alpha}O=0$, and $Q^2 K = \bar{Q}^2 K = 0$, so 
  $\tr(O\,K)$ is annihilated only by $\bar{Q}^2$.}.
One possible outcome, for instance, could be a comparison with  
entropy counting of the recently constructed $AdS_5$ SUSY black holes
\footnote{Notice that these BPS black holes are constructed in
  $5$-dimensional gauged supergravity  
  with $U(1)^k$ gauge group, so can in principle be uplifted to various
  $AdS_5 \times X^5$ solutions.}.

In $\cN=4$ SYM an interesting problem is whether there is a change in the
number of BPS operators changing the coupling. At zero coupling one
expects more BPS states. Here, there is a precisely analogous
question. It was shown in \cite{Benvenuti:2005cz} that, 
in the moduli space of
SCFTs corresponding to a given quiver gauge theory, there is a special
point with enhanced chiral ring\footnote{For $\cN=4$ and orbifolds
  thereof 
  this special point is the free theory, for the conifold it
  corresponds to having vanishing superpotential. For generic quivers
  we have only one term of the superpotential vanishing.}.
It was
observed in \cite{Benvenuti:2005cz} that at this special point, 
the growth of
the number of single-trace, $N=\infty$, BPS mesons is exponential
instead of that quadratic (as is the case on generic points of the moduli
space of SCFTs). It would be interesting to study further this
mechanism, that could lead at finite temperature to phase
transitions.

On the gravity/string side of AdS/CFT, we should also find the same
partition functions. For single-trace operators the result is
well-known. The interesting case is multi-trace at finite $N$. One way
to reproduce the $g_N$ should be counting Giant Gravitons (GG's) 
in $AdS_5 \times X^5$. There are studies of GG's in $AdS_5
\times S^5$ and $AdS_5 \times T^{1,1}$ \cite{GG}.
In the case of $S^5$, one considers the classical moduli
space of GG's and is led to study $N$ classical non-interacting
particles, whose single particle phase space
is $\IC^3$. Quantizing the multi particle phase space one
gets $N$ bosons whose single particle states are the integer
points of the toric cone of $\IC^3$. The partition function is
precisely the finite $N$ partition function of 1/8-BPS ops in $\cN=4$
SYM. In generic toric quivers, for instance, one should find the
result of the end of \sref{s:finiteN}:
 \[
 \sum_N g_N(t_i) \nu^N = \prod_{n,m,r  \in C} \frac{1}{(1 - \nu
 t_1^n  t_2^m  t_3^r)}
 \]
A different approach is \cite{Mandal:2006tk}, where they consider ``dual
GG's'', i.e., D3-brane wrapping an $S^3$ inside $AdS_5$, and moving along a
trajectory in $X^5$. For generic Sasaki-Einstein manifolds $X^5$, we
conjecture that these states are BPS if and only if the trajectory is
a BPS geodesic. We already know that single-trace BPS mesons are the
quantization of such BPS geodesics (see \cite{Benvenuti:2005cz}),  
so the final
result should be a simple outcome of the use of the Plethystic
Exponential on the gravity side.

{\vskip 1in}

\section*{Acknowledgements}
We heartily acknowledge James Spraks, Dario Martelli and Davide
Gaiotto for wonderful communications.
S.~B.~is indebted to Antonello Scardicchio
for discussions and the ``Fondazione Angelo della Riccia'' for
financial support. 
B.~F.~is obliged to the Marie Curie Research
Training Network under contract number MRTN-CT-2004-005104 and extends
his thanks to Merton College, Oxford for warm reception at the final
stages of the draft. 
A.~H.~is grateful to Alastair King,
Marcos Mari\~{n}o and Yuji Okawa for
enlightening conversations, to Philip Candelas and Xenia
de la Ossa for kind hospitality at the University of Oxford and Jerome
Gauntlett, at Imperial College, London, as well as the Technion and
Tel Aviv University as the manuscript was being completed.
Y.~H.~H. bows to the gracious patronage, via the FitzJames Fellowship,
of Merton College, Oxford.

\newpage

%%------------
%=+++++++++++++++++++++++++++++++++++++++

\end{document}